\documentclass[aps,amssymb,amsmath,prd,twocolumn,
showpacs,preprintnumbers,superscriptaddress,nofootinbib,floatfix]{revtex4-1}
\usepackage{graphicx}
\usepackage{bm}
\usepackage{mathrsfs}
\usepackage{latexsym}
\usepackage{float}
\usepackage{color}
\usepackage[normalem]{ulem} 
\usepackage{dcolumn}
\usepackage[colorlinks=true,citecolor=blue,urlcolor=blue]{hyperref}
\usepackage[usenames,dvipsnames]{xcolor}
\usepackage[caption=false]{subfig}
\usepackage{slashed}

\newcommand\Eqn[1]{Eq.~(\ref{#1})}  

\newcommand{\p}{\partial}
\newcommand{\nn}{\nonumber}
\newcommand{\be}{\begin{equation}}
\newcommand{\ee}{\end{equation}}
\newcommand{\beq}{\begin{eqnarray}}
\newcommand{\eeq}{\end{eqnarray}}

\newcommand{\nb}{n_{\rm B}}
\newcommand{\nsat}{n_{\rm sat}}

\newcommand{\ep}{\varepsilon}

\newcommand{\csqeq}{c_{\rm eq}^2}
\newcommand{\csqad}{c_{\rm ad}^2}
\newcommand{\ceq}{c_{\rm eq}}
\newcommand{\cad}{c_{\rm ad}}

\newcommand{\zhao}[1]{\textcolor{magenta}{TZ: #1}}

\newcommand{\cons}[1]{\textcolor{orange}{C: #1}}

\begin{document}
\title{Quasi-normal g-modes of neutron stars with quarks} 

\author{Tianqi Zhao}
\email{zhaot@ohio.edu}
\affiliation{Department of Physics and Astronomy, Ohio University,
Athens, OH~45701, USA}

\author{Constantinos Constantinou}
\email{cconstantinou@ectstar.eu}
\affiliation{INFN-TIFPA, Trento Institute of Fundamental Physics and Applications, Povo, 38123 TN, Italy}
\affiliation{European Centre for Theoretical Studies in Nuclear Physics and Related Areas, Villazzano, 38123 TN, Italy}

\author{Prashanth Jaikumar}
\email{prashanth.jaikumar@csulb.edu}
\affiliation{Department of Physics and Astronomy, California State University Long Beach, Long Beach, CA~90840, USA}

\author{Madappa Prakash}
\email{prakash@ohio.edu}
\affiliation{Department of Physics and Astronomy, Ohio University,
Athens, OH~45701, USA}

\date{\today}

\begin{abstract}
Quasi-normal oscillation modes of neutron stars provide a means to probe their interior composition using gravitational wave astronomy. We compute the frequencies and damping times of composition-dependent core $g$-modes of neutron stars containing quark matter employing linearized perturbative equations of general relativity. We find that ignoring background metric perturbations due to the oscillating fluid, as in the Cowling approximation, underestimates the $g$-mode frequency by up to 10\% for higher mass stars, depending on the parameters of the nuclear equation of state and how the mixed phase is constructed. The $g$-mode frequencies are well-described by a linear scaling with the central lepton (or combined lepton and quark) fraction for nucleonic (hybrid) stars. Our findings suggest that neutron stars with and without quarks are manifestly different with regards to their quasi-normal $g$-mode spectrum, and may thus be distinguished from one another in future observations of gravitational waves from merging neutron stars.
\end{abstract}

\maketitle


\section{Introduction}
\label{sec:intro}

Neutron Stars (NSs) are natural laboratories to study the behaviour of matter under extreme conditions of density, rotation and magnetic fields~\cite{glendenning2012compact, LP-review}. They are observed across many wavelengths of the electromagnetic spectrum, from radio waves to X-rays and gamma-rays, using a variety of instruments and telescopes~\cite{LP-review2}. Recent observations of merging NSs via gravitational waves have ushered in a multi-messenger characteristic to research in this area (for recent reviews, 
see \cite{metzger2017kilonovae,bailes2021gravitational}). Pioneering measurements of pulse modulation profiles by the recently launched NICER (Neutron Star Interior Composition Explorer) mission is advancing the goal of constraining the equation of state of dense matter by setting precise limits ($\approx 5\%)$ on the radius of a neutron star~\cite{Riley-nicer,Miller-nicer}.

However, the interior composition of the NS core is likely to remain uncertain if only global and static properties of the star, like mass and radius, are measured. Even with few \% precision in these quantities, one cannot distinguish between the many forms and ways in which exotic matter can appear in the core of neutron stars~\cite{Baym:2017whm,Han:2019bub,2019ApJ...887..151W,Kapusta:2021ney}. Definitive signatures of such exotic matter in static NS observables is elusive, but new frontiers in multi-messenger astronomy, such as gravitational waves, can provide new insight. For example, when two neutron stars merge, the tidal forces from one component NS on the other can excite secular quasi-normal modes (QNMs) that affect the phase of the gravitational waveform~\cite{PhysRevD.101.103009,Yu17}.  Characteristics of some of these QNMs, in particular, the core $g$-mode~\cite{RG,RG2,jaikumar2021g,2021PhRvD.104.l23032C} depend strongly on the composition of the star. Detecting these QNMs in mergers or isolated NSs remains an observational challenge, but in principle, this offers a method to constrain the composition.

The study of QNMs originated in black holes with the work of Regge \& Wheeler~\cite{PhysRev.108.1063}, Vishveshwara~\cite{vishveshwara_1970} and Zerilli~\cite{PhysRevLett.24.737}. The application to neutron stars was begun by Thorne and Campolattaro~\cite{1967ApJ...149..591T}, followed by numerous other works (for a review, see ~\cite{1999LRR.....2....2K} and relevant references therein). Normal modes of neutron stars are traditionally categorized by the restoring force that tries to bring the perturbed star back to equilibrium. Not all of these modes couple to gravitational waves, but all of them are typically subject to dissipation, hence they are regarded as ``quasi"-normal modes.

Our focus in this work is on the $g$-mode since it is an example of a secular~\footnote{ The term secular refers to the fact that some of these modes are long-lived, as the dissipation is small.} QNM that appears to be strongly correlated with the composition of the core, displaying a remarkable sensitivity to the onset of new species of particles~\cite{jaikumar2021g,2021PhRvD.104.l23032C}. To be clear, these are different from the oft-studied discontinuity $g$-modes~\cite{Tonetto:2020bie,Sotani:2001bb} which arise from a sharp change in the density in the interior, as for example at the boundary of a Maxwell-type first-order transition~\cite{miniutti2003non, lau2021probing, zhao2022universal} or the core-crust boundary~\cite{Ranea-Sandoval:2018bgu}. The composition dependent $g$-modes studied here arise in the NS core and their characteristic frequencies lie within the sensitivity range of current generation of gravitational wave (GW) detectors. They could be excited by strong tidal effects during the inspiral phase of NS mergers~\cite{Pratten:2019sed}. We also note that while studies of the $g$-mode abound in the literature, several choose to employ the Cowling approximation~\cite{Pereira:2017rmp,Kantor:2014lja,McD83}. In fact, the composition $g$-modes pioneered in the work of Reisenegger and Goldreich~\cite{RG,RG2} used Newtonian gravity for simplicity, and subsequent works ~\cite{Kantor:2014lja,Dommes:2015wul,2021arXiv211000324K} used the relativistic Cowling approximation in lieu of a fully general relativistic treatment. There are no extant studies of composition $g$-modes for realistic neutron or hybrid stars that employ general relativity (GR). We take this opportunity to ``fill the gap" in the literature, as well as to shed new light on how gravitational wave astronomy could play an important role in uncovering the role of quarks or other forms of exotic matter in neutron stars.

In recent works~\cite{jaikumar2021g,2021PhRvD.104.l23032C}, we investigated the role of composition gradients, including phase transitions to quark matter, on the sound speeds and $g$-modes in neutron stars. We found a rise in the frequency of the $g$-mode at the onset of quark matter that was most pronounced in the case of a Gibbs mixed phase~\cite{jaikumar2021g} and less so in the case of a crossover~\cite{2021PhRvD.104.l23032C}. These results were obtained in the relativistic Cowling approximation (which neglects perturbations of the background metric). Completing the analysis and confirming that this effect is robust, we present the results of calculations of such core $g$-modes of hybrid stars in a fully general relativistic framework. We compare our results in GR to those from the Cowling approximation, finding generally good agreement as expected. We also compute the damping time of core $g$-modes due to gravitational wave emission, which is only made possible by the general relativistic prescription~\footnote{We do not explicitly report results for damping times in this work. It suffices to mention that this damping time is found to be extremely long compared to merger timescales for non-rotating stars. However, $g$-modes of neutron stars that rotate sufficiently rapidly can have much shorter damping times~\cite{Lai:1998yc}, making them unstable to gravitational wave emission. Therefore such modes remain relevant to merging neutron stars with high spins or rapidly rotating nascent neutron stars in the post-merger phase~\cite{Kruger2021}.}.

This paper is organized as follows. Section \ref{sec:GRR} is a review of the linearized equations in GR that describe polar oscillation modes of an ideal self-gravitating fluid. This recap largely follows the presentation by Lindblom and Detweiler~\cite{Detweiler83,Detweiler85}, and establishes our working equations up front, and their reduction to the Cowling approximation. Section \ref{sec:NUC} describes the specific EOS employed to treat the phase transition, along with a discussion of the relevant parameters that affect basic neutron star properties (mass,radius,composition). In Sec. \ref{results}, we present a comparison of our results in GR against the Cowling approximation, as well as some scaling relations. We also discuss the relevance of our results to compact stars and gravitational wave searches for their quasi-normal oscillation modes in this section, and summarize our conclusions in Sec. \ref{sec:concs}.

\section{Asteroseismology in General Relativity} 
\label{sec:GRR}

The asteroseismology of compact objects like neutron stars and black holes requires a general relativistic framework to address the stability of the oscillations and emission of gravitational waves~\cite{1991RSPSA.432..247C}. The main oscillation modes and their implications for asteroseismology are reviewed in~\cite{andersson1998towards,lasky_2015,Promises}. For the $f$-mode of neutron stars, which is sensitive to compactness~\cite{andersson1998towards} or moment of inertia~\cite{lau2010inferring} and static tidal polarizability~\cite{chan2014multipolar,2021arXiv210908145S}, but not to composition~\footnote{An additional scaling with compactness arises since the modes are observed at ``infinity" and must be corrected for the gravitational redshift at the star's surface.}, post-Newtonian formulae can provide reasonable estimates of its frequency and damping time~\cite{andersson1998towards}. These estimates hold up well in a general relativistic calculation and for different microscopic EOS. As pointed out recently~\cite{universe7040097}, the existence of universal (EOS-independent) scaling relations for the $f$-mode frequency and damping time in general relativity are not unexpected, given that analytically solvable models in general relativity, such as the Tolman VII solution, have density profiles in accordance with a wide range of EOS~\cite{2001ApJ...550..426L}.

Whether $g$-modes, which arise from compositional differences, exhibit any sort of scaling with global stellar or material properties in general relativity is an open question. Another reason to study $g$-modes in GR is that they can be dynamically excited by tidal forces in a binary merger~\cite{Lai:1993di}, providing compositional information in the early gravitational wave signal that cannot be gleaned from the static tidal deformability/Love number. This compositional component in the dynamical tide is expected to be small compared to the $f$-mode~\cite{PhysRevD.101.083001}, but may be teased out even at the few percent level by future detectors such as the Einstein telescope or the Cosmic Explorer~\cite{universe7040097}. As an aside, perturbations of classical gravitational backgrounds in GR also serve as a tool for computing transport properties of strongly coupled quantum field theories via the gauge-gravity holographic duality with applications to the quark-gluon plasma~\cite{2009CQGra..26p3001B}.

\subsection{Non-radial oscillations in General Relativity}

Thorne {\it et~al.} first studied NS oscillations coupled with gravitational radiation  \cite{thorne1967non}. Oscillations of NSs are expected to be linear scalar variations of pressure and density. Since scalar variations in spherical harmonics are of even parity, only even-parity perturbations of the Regge-Wheeler metric are considered:
\begin{equation}
\begin{split}
ds^2 = - e^{\nu(r)} [1+r^l H_0(r)e^{i\omega t} Y_{lm}(\phi,\theta)]c^2 dt^2\\
+e^{\lambda(r)} [1-r^l H_0(r)e^{i\omega t} Y_{lm}(\phi,\theta)]dr^2 \\
+ [1-r^l K(r)e^{i\omega t}Y_{lm}(\phi,\theta)]r^2 d\Omega^2\\
-2i\omega r^{l+1}H_1(r)e^{i\omega t}Y_{lm}(\phi,\theta) dt~dr\,,
\end{split}
\end{equation}
where 
\begin{equation}
e^{\lambda(r)}=\frac{1}{1-\frac{2 G m(r)}{c^{2} r}}
\end{equation}
and 
\begin{align}
e^{\nu(r)} = \exp \bigg(-\frac{2 G}{c^{2}} \int_{0}^{r}\left\{\frac{\left[m(r^{\prime})+\frac{4 \pi p(r^{\prime}) r^{\prime 3}}{c^{2}}\right]}{r^{\prime}\left[r^{\prime}-\frac{2 m(r^{\prime}) G}{c^{2}}\right]}\right\} d r^{\prime}\bigg){\rm e}^{\nu_0},
\label{eq:static_metric_nu}
\end{align}
where $m(r^\prime)$ is the enclosed mass of the star at $r^\prime$ and $p$ is the pressure at $r$. The functions
$H_0$, $H_1$, and $K$ are radial perturbations of the metric and the angular part is contained in the spherical harmonics $Y_{lm}$ with $l$ denoting the orbital angular momentum number and $m$ the azimuthal number. The quantity $\omega$ is a complex oscillation frequency; its real component is the oscillation frequency and its imaginary component is the inverse of the damping time (if it is positive). We can compute both in the GR framework.  The perturbations of the metric inside the star and outside the star must match at the stellar surface. The factor ${\rm e}^{\nu_0}$~\cite{1991RSPSA.432..247C} accounts for a matching of the exterior and interior unperturbed metric.

Perturbations of the fluid inside the star are described by the Lagrangian displacement vector
\begin{eqnarray}
\xi^r &=& r^{l-1}e^{-\frac{\lambda}{2}}W Y^l_m e^{i\omega t} \label{eq:xi_radial}\\
\xi^\theta &=& -r^{l-2} V \partial_\theta Y_m^l e^{i\omega t}\\
\xi^\phi &=& -\frac{r^{l-2}}{ \sin^{2}\theta} V\partial_\phi Y_m^l e^{i\omega t} \,,
\end{eqnarray}
which defines the perturbation amplitudes $W$ and $V$, with a dimension of $[R]^{2-l}$ where $R$ is the radius of the star. \\

Perturbations of a spherical star have four degrees of freedom; three coming from the metric perturbations, which will be reduced by one applying Einstein's equation, $\delta G^{01}=8\pi \delta T^{01}$, and two coming from the fluid perturbations.  An additional function $X$, related to Lagrangian pressure variations, in addition to $H_0$, $H_1$ and $K$ is defined as 
\begin{equation}
\Delta p = -r^{l}e^{-\frac{\nu}{2}}X Y_m^l e^{i\omega t}\,.
\end{equation}

In order to avoid a singularity in the fourth-order ODEs governing NS oscillation for some frequency range, Lindblom {\it et~al.}  \cite{Lindblom:1983,Detweiler:1985}   
choose the four degrees of freedom to be $H_1$, $K$, $W$, and $X$.  Evaluating the two remaining functions $H_0$ and $V$ in terms of them yields
\begin{eqnarray}
H_0&=&\left \{8\pi r^2 e^{-\nu/2}X-\left[(n+1)\textrm{Q}-\omega^2r^2 e^{-(\nu+\lambda)}\right]H_1 \nonumber \right. \\
&+& \left. \left [ n-\omega^2 r^2 e^{-\nu}-e^\lambda \textrm{Q}(\textrm{Q}-e^{-\lambda})  \right ] K \right \} (2b+n+\textrm{Q})^{-1}\label{eq:H0_def_in},\nonumber \\ \\
V&=&\left[\frac{X}{\varepsilon+p}-\frac{\textrm{Q}}{r^2}e^{(\nu+\lambda)/2}W-e^{\nu/2}\frac{H_0}{2}\right]\frac{e^{\nu/2}}{\omega^2}\label{eq:V_def_in} \,,
\end{eqnarray}
where $n=(l-1)(l+2)/2$, $b=Gm/(rc^2)$, $\textrm{Q}=b+4\pi G r^2 p/c^4$ and $\varepsilon$ is the local energy density. 
By expanding Einstein's equation to first-order, the homogeneous linear differential equations for $H_1$, $K$, $W$ and $X$ are  \cite{Detweiler:1985},
\begin{eqnarray}
r\frac{dH_1}{dr}&=&-[l+1+2b e^\lambda+4\pi r^2 e^\lambda(p-\varepsilon)] H_1\nonumber\\
&+& e^\lambda[H_0+K-16\pi(\varepsilon+p)V]\,, \label{eq:ODE_DL1} \\
r\frac{dK}{dr}&=& H_0+(n+1)H_1 \nonumber \\
&+&[e^\lambda \textrm{Q}-l-1]K-8\pi(\varepsilon+p)e^{\lambda/2}W \,, \label{eq:ODE_DL2}\\
r\frac{dW}{dr}&=&-(l+1)[W+le^\frac{\lambda}{2}V] 
\nonumber \\
&+&r^2 e^{\lambda/2}\left[\frac{e^{-\nu/2}X}{ (\varepsilon+p) c_{\rm ad}^2}+\frac{H_0}{2}+K\right., \label{eq:ODE_DL3}\\
r\frac{dX}{dr}&=& \left.-lX+\frac{(\varepsilon+p)e^{\nu/2}}{2} \right.\nonumber \\
&&\left\{ (3e^\lambda\textrm{Q}-1)K
-\frac{4(n+1)e^\lambda\textrm{Q}}{r^2}V\right.\nonumber  \\
&& \left. +(1-e^\lambda \textrm{Q})H_0+(r^2\omega^2e^{-\nu}+n+1)H_1 \right. \nonumber\\
&& -\left[8\pi(\varepsilon+p)e^{\lambda/2}+2\omega^2 e^{\lambda/2-\nu} \right.  \nonumber \\ 
&& \left. \left. -r^2\frac{d}{dr} \left(\frac{e^{-{\lambda/2}}}{r^2} \frac{d\nu}{dr} \right)\right]W \right\} \,, \label{eq:ODE_DL4} 
\end{eqnarray}
where $c_{\rm ad}^2=\partial p/\partial\varepsilon$ is the adiabatic sound speed of NS matter under oscillations, hence the $\partial$ here denotes a derivative taken at fixed composition, i.e., assuming all weak reactions are too slow compared to the oscillation timescale. This is different from the equilibrium sound speed $c_{\rm eq}^2=dp/d\varepsilon$ \cite{wei2020lifting,jaikumar2021g} where weak equilibrium is assumed to be restored effectively instantaneously.

The boundary conditions for the perturbation functions at the center of the star $r=0$ are
\begin{eqnarray}
W(0)&=&1\\
X(0)&=&(\varepsilon_0+p_0)e^{\nu_0/2} \nonumber \\
&&\hspace*{-.2cm} \left \{ \left[ \frac{4\pi}{3}(\varepsilon_0+3p_0)-\frac{\omega^2}{l} e^{-\nu_0}\right]W(0)+\frac{K(0)}{2}\right \} \nonumber \\
\label{eq:BC_X0} \\
H_1(0)&=&\frac{lK(0)+8\pi(\varepsilon_0+p_0)W(0)}{n+1}\\
X(R)&=&0 \,,
\end{eqnarray}
where the last boundary condition is 
obtained
by solving the two trial solutions with $K(0)=\pm(\varepsilon_0+p_0)$ and then linearly constructing the correct solution satisfying the boundary condition $X(r=R)=0$ (no pressure variations at the surface). Note that $H_0(0)=K(0)$ by construction.

\subsection{Relativistic Cowling Approximation}

In Newtonian theory of stellar pulsations, when 
the perturbation of the gravity field due to matter perturbation is neglected, this approximation is known as the Cowling approximation~\cite{McD83}, and the resulting equations for the fluid perturbations are considerably simpler. Similarly, in the case of the relativistic theory, the perturbation of the GR metric is often neglected as well, leading to the {\it relativistic} Cowling approximation. The relativistic Cowling equations are obtained by setting $H_0=H_1=K=0$ in Eq. (\ref{eq:V_def_in}), Eq. (\ref{eq:ODE_DL3}) and Eq. (\ref{eq:ODE_DL4}), and furthermore, omitting the term $-4\pi(\varepsilon+p)^2e^{(\nu+\lambda)/2}W$ in Eq. (\ref{eq:ODE_DL4_cowling}), leading to 
\begin{eqnarray}
\label{eq:X_def_in_cowling}
V&=&\left[\frac{X}{\varepsilon+p}-\frac{d
\Phi}{d \ln r}e^{(\nu-\lambda)/2}\frac{W}{r^2}\right]\frac{e^{\nu/2}}{\omega^2},\\
\frac{dW}{d\ln{r}}&=&-(l+1)\left[W+le^\frac{\lambda}{2}V\right]+r^2 e^{\lambda/2}\left[\frac{e^{-\nu/2}X}{ (\varepsilon+p) c_{ad}^2 }\right],\label{eq:ODE_DL3_cowling}\\
\frac{dX}{d\ln{r}}&=&-lX+(\varepsilon+p)e^{\nu/2}
\left\{
-2(n+1)\frac{d\Phi}{d\ln r}\frac{V}{r^2}\right.\nonumber\\
&&\hspace*{-.5cm}\left. -\left[\omega^2 e^{\lambda/2-\nu}-r^2\frac{d}{dr} \left(\frac{e^{-{\lambda/2}}}{r^2} \frac{d\Phi}{dr} \right)\right]W \right\}\,, \label{eq:ODE_DL4_cowling}
\end{eqnarray}
where $\Phi=2\nu$ as in Eq. (\ref{eq:static_metric_nu}). The reason for omitting the term $-4\pi(\varepsilon+p)^2e^{(\nu+\lambda)/2}W$ in Eq. (\ref{eq:ODE_DL4_cowling}) is that it follows directly from the last term on the right hand side of Eq. (\ref{eq:ODE_DL2}), wherein an application of the Cowling approximation implies that this term should vanish for self-consistency.\\

 The boundary conditions for Eq.  (\ref{eq:ODE_DL3_cowling}) and Eq. (\ref{eq:ODE_DL4_cowling}) are obtained by setting $K=0$ in Eq. (\ref{eq:BC_X0}):
\begin{eqnarray}
X(0)&=&(\varepsilon_0+p_0)e^{\nu_0/2} \left[ \frac{4\pi}{3}(\varepsilon_0+3p_0) \right. \nonumber \\
&& \hspace*{-.4cm} \left.-\frac{\omega^2}{l} e^{-\nu_0}\right]W(0)\\
X(R)&=&0 \,,
\end{eqnarray}
where the first condition is an initial value condition for the ODE system Eq. (\ref{eq:ODE_DL3_cowling}) and Eq. (\ref{eq:ODE_DL4_cowling}), and the second one is the boundary condition that determines the eigenvalue of the oscillation frequency. Unlike the GR equations, where one needs to determine the appropriate linear combination of two trial functions at the origin that satisfy the surface boundary condition, the initial values in the Cowling approximation are uniquely fixed, which speeds up the computational time by at least a factor of two. Obviously, reducing four ODEs to two ODEs is another big advantage of the Cowling approximation.

Although the set of ODEs, Eqs.(\ref{eq:ODE_DL3_cowling}), (\ref{eq:ODE_DL4_cowling}) represent the equations obtained in the Cowling approximation, neither they nor the boundary conditions above are generally used in this form for the purpose of calculating the oscillation frequencies. The variable $X$ that appears in the GR formalism can, in the Cowling approximation, be eliminated in favor of $W,V$ by inserting Eq. (\ref{eq:X_def_in_cowling}) into Eq. (\ref{eq:ODE_DL3_cowling}) and Eq. (\ref{eq:ODE_DL4_cowling}) to obtain
\begin{eqnarray}
\frac{dW}{d\ln{r}}&& =-(l+1)\left[W+le^\frac{\lambda}{2}V\right]
\nonumber \\
&&+\frac{(\omega r)^2 e^{\lambda/2-\nu}}{c_{ad}^2}\left [V+ \frac{e^{\nu-\lambda/2}}{(\omega r)^2}\frac{d\Phi}{d\ln r}W \right]\,,\\ \label{eq:ODE_DL3_cowling_VW}
\frac{dV}{d\ln{r}}&& = (2\frac{d\Phi}{d\ln r}-l)V \nonumber \\
&& +e^{\lambda/2}W+\Delta(c^{-2})\frac{d\Phi}{d\ln r}\left [V+ \frac{e^{\nu-\lambda/2}}{(\omega r)^2}\frac{d\Phi}{d\ln r}W \right] \,, 
\label{eq:ODE_DL4_cowling_VW} 
\end{eqnarray}
where $\Delta(c^{-2})=\frac{1}{c_{\rm eq}^2}-\frac{1}{c_{\rm ad}^2}$. The suitable boundary conditions for these equations is discussed below.

The above two ODEs can be simplified further by defining $U=-e^{-\nu}V$,
\begin{eqnarray}
\label{eq:ODE_DL3_cowling_UW}
\frac{dW}{d\ln r} && = -(l+1)\left[W-l e^{\nu +\lambda/2} U\right] \nonumber \\
&& -\frac{e^{\lambda/2}(\omega r)^2}{c_{ad}^2}\left[U-\frac{d\Phi}{d \ln r} \frac{e^{-\lambda/2}}{(\omega r)^2} W\right ]\,,\\
\frac{dU}{d\ln r} && = e^{\lambda/2-\nu}\left[W -le^{\nu-\lambda/2}U\right] \nonumber \\
&& + \Delta(c^{-2}) \frac{d\Phi}{d \ln r} \left [ U -\frac{d\Phi}{d \ln r} \frac{e^{-\lambda/2}}{(\omega r)^2} W\right ]\,,\label{eq:ODE_DL4_cowling_UW}
\end{eqnarray}
where $W=e^{\lambda/2} r^{1-l} \xi^r$ and $U=r^{-l} \omega^{-2} \delta p/(\varepsilon+p)$~\footnote{As a crosscheck on the limiting procedure, Eqs.(\ref{eq:ODE_DL3_cowling_UW}) and (\ref{eq:ODE_DL4_cowling_UW}) are identical to the Cowling approximation Eqs.(79) in~\cite{jaikumar2021g} with the replacements $W\rightarrow {\cal U}/r^{l+1}$, $U\rightarrow {\cal V}/(\omega^2 r^l)$, where ${\cal U},{\cal V}$ indicate the fluid variables used in Eqs.(79) of~\cite{jaikumar2021g}.}, $\xi^r$ are proper radial Lagrangian displacements defined in Eq. (\ref{eq:xi_radial}) and $\delta P$ is the Eulerian perturbation of pressure, which is related to the the Lagrangian perturbation by $\Delta P=\delta P-(\varepsilon+p) \frac{d\Phi}{dr} \xi^r$. This form is particularly advantageous because the boundary conditions are explicitly collected in square brackets. The first square brackets of Eq. (\ref{eq:ODE_DL3_cowling_UW}) and Eq. (\ref{eq:ODE_DL4_cowling_UW}) vanish at $r=0$, whereas the second square brackets are proportional to the Lagrangian pressure variation, which vanishes at $r=R$. 
Explicitly, the boundary conditions can be written as
\begin{eqnarray}
\left.\frac{W}{U}\right|_{r=0}&=&l e^{\nu_c}=l e^{2\Phi_c} \\
\left.\frac{W}{U}\right|_{p=0}&=& \frac{e^{\lambda/2}(\omega r)^2}{\frac{d\Phi}{d \ln r}}
= \frac{(\omega r)^2\sqrt{1-2b}}{b+4\pi r^2p}|_{p=0}\nonumber\\
&=&(\omega R)^2  \frac{c^2R}{GM}\sqrt{1-\frac{2GM}{c^2R}} \,,
\end{eqnarray}
which determines the eigenmode frequency of the oscillation. Note that only the ratio between W and U matters instead of the individual magnitudes of W and U, as the corresponding eigenmode can have an arbitrary amplitude. Thus, we simply take $W(r=0)=1$ and $U(r=0)= e^{-2\Phi_c}/l$. Because the two ODEs for W and U are homogeneous ODEs, we can solve the ODE of $W/U$ directly:
\begin{eqnarray}
\frac{d(W/U)}{d\ln r} && = -(l+1)\left[\frac{W}{U}-l e^{\nu +\lambda/2}\right] \nonumber \\
&& -e^{\lambda/2-\nu}\frac{W}{U}\left[\frac{W}{U} - le^{\nu-\lambda/2}\right] \nonumber\\
&&-\left [\frac{e^{\lambda/2}(\omega r)^2}{c_{ad}^2}+\Delta(c^{-2}) \frac{d\Phi}{d \ln r} \frac{W}{U}\right ] \nonumber \\ 
&&\times \left[1-\frac{d\Phi}{d \ln r} \frac{e^{-\lambda/2}}{(\omega r)^2} \frac{W}{U}\right ]
\end{eqnarray}
In this form, we only need to solve one ODE instead of two ODEs, and the boundary conditions are more straight forward to implement.

\section{Equations of state with and without quarks} 
\label{sec:NUC}

The discussion in the previous section shows that the pressure $p$, energy density $\varepsilon$, and their derivatives $c_{\rm eq}^2=dp/d\varepsilon$ and $c_{\rm ad}^2=\partial p/\partial\varepsilon$ inside the NS feature prominently in determining $g$-modes both in GR and in the Cowling approximation.  In what follows, we briefly describe the equations of state (EOSs), the relation between $p$ and $\varepsilon$, used in this work, both for the case of nucleons- and leptons-only matter and that including quarks. In the latter case, we consider quarks appearing according to the Gibbs construction as well as when there is a smooth crossover. %
The case of discontinuous $g$-modes~\cite{finn1987g,mcdermott1990density} encountered in the case when Maxwell construction is used to treat first-order phase transitions is considered in a separate work \cite{zhao2022universal}.
%

\subsection{Nucleonic matter EOS}
We describe nucleonic matter using the Zhao-Lattimer (ZL) \cite{Zhao:2020dvu} model. We adjust its parameters such that the nuclear saturation density $\nsat = 0.16~{\rm fm}^{-3}$, the binding energy $E_{\rm sat}$ = -16 MeV, the compression modulus $K_{\rm sat}$ = 230 MeV, and the symmetry energy $S_v$ = 31 MeV. The high-density behavior is controlled by varying the slope of the symmetry energy parameter, $L$, within the range 40 - 70 MeV, and a power-law index $\gamma_1$ fixed at a value of 2.

For a nucleonic system in which baryon number  conservation,  charge  neutrality  and $\beta$-equilibrium  have not been imposed, the independent variables are the baryon density $\nb$ and the individual nucleon fractions $y_n$, $y_p$. The total energy density of such a system, as given by the ZL functional, is 
\beq
\ep_H &=& \frac{1}{8\pi^2\hbar^3}\sum_{h=n,p}\left\{k_{Fh}(k_{Fh}^2 + m_H^2)^{1/2}(2k_{Fh}^2 + m_H^2)\right.  \nn \\
 &-& \left. m_H^4\ln\left[\frac{k_{Fh}+(k_{Fh}^2 + m_H^2)^{1/2}}{m_H}\right]\right\} \nn \\
           &+& 4 \nb^2 y_n y_p \left\{\frac{a_0}{\nsat} 
            +\frac{b_0}{\nsat^{\gamma}} [\nb(y_n + y_p)]^{\gamma - 1}\right\}  \nn \\
            &+& \nb^2 (y_n - y_p)^2\left\{\frac{a_1}{\nsat} 
            + \frac{b_1}{\nsat^{\gamma_1}}[\nb(y_n + y_p)]^{\gamma_1-1}\right\}  . \nn \\
\eeq
Here $m_H=939.5$ MeV is the common nucleon mass, and $k_{Fh} = (3\pi^2\hbar^3\nb y_h)^{1/3}$ is the Fermi momentum of nucleon species $h$. Note that, in the present context, ``$H$'' and ``$h$'' denote nucleons. The parameters $a_0, b_0$ and $\gamma$ refer to isospin-symmetric matter, whereas $a_1, b_1$ and $\gamma_1$ to isospin-asymmetric matter.

We get the chemical potentials from (see \cite{2021PhRvD.104.l23032C} for explicit expressions)
\be
\mu_{h_1} = \frac{1}{n_B}\left.\frac{\partial \varepsilon_H}{\partial y_{h_1}}\right|_{n_B, y_{h_2}}~,
\ee
the pressure from the thermodynamic identity 
\be
p_H = \nb\sum_{h=n,p}\mu_h y_h - \ep_H ~,
\ee 
and the equilibrium speed of sound from
\be
\left(\frac{\ceq}{c}\right)^2 =  \frac {d p_H}{d  \ep_H} ~.
\ee
The adiabatic speed of sound is obtained from the partial derivatives of the pressure and the total energy density with respect to baryon density with all particle fractions fixed
\be
\left(\frac{\cad}{c}\right)^2 = \left.\frac{\p p_H}{\p \nb}\right|_{y_h}
                           \left(\left.\frac{\p \ep_H}{\p \nb}\right|_{y_h}\right)^{-1}~.
\ee

\subsection{Quark matter EOS}

For the calculation of the quark EOS, we use the vMIT bag model~\cite{Gomes:2018eiv, Klahn:2015mfa} whose Lagrangian density is given by
\be
\mathcal{L}=\sum_{q=u,d,s}\left[\bar{\psi}_{q}\left(i \slashed{\p} -m_{q}-B\right) \psi_{q}+\mathcal{L}_{\mathrm{vec}}\right] \Theta \,,
\ee
where $\mathcal{L}_{\mathrm{vec}}$ describes repulsive interactions between quarks of mass $m_{q}$ confined within a bag (denoted by the $\Theta$ function):
\be
\mathcal{L}_{\text {vec }}=-G_{v} \sum_{q} \bar{\psi} \gamma_{\mu} V^{\mu} \psi+\left(m_{V}^{2} / 2\right) V_{\mu} V^{\mu} \,.
\ee
$B$ is a constant reflecting the cost of confining the quarks inside the bag, and the $m_{q}$ are the current quark masses (here, $m_u = 5$ MeV, $m_d = 7$ MeV, and $m_s = 150$ MeV). \\

The energy density, chemical potentials, and pressure, corresponding to the above Lagrangian (before the application baryon number conservation, charge neutrality, and chemical equilibrium) are
\beq
\ep_Q &=& \sum_{q=u,d,s}\ep_q + \frac{1}{2}a~\hbar~[\nb(y_u+y_d+y_s)]^2 
                + \frac{B}{\hbar^3}   \\
\ep_q &=& \frac{3}{8\pi^2\hbar^3}\left\{k_{Fq} (k_{Fq}^2 + m_q^2)^{1/2}(2 k_{Fq}^2 + m_q^2) 
\right.      \nn \\      
        &-& \left. m_q^4\ln\left[\frac{k_{Fq}+(k_{Fq}^2 + m_q^2)^{1/2}}{m_q}\right]\right\}   \\
\mu_q &=& (k_{Fq}^2 + m_q^2)^{1/2} + a~\hbar~\nb (y_u + y_d + y_s)        \\
p_Q &=& \nb \sum_{q=u,d,s}  \mu_q y_q - \ep_Q
\eeq
where $a \equiv (G_v/m_V)^2$ and $k_{Fq} = (\pi^2 \hbar^3 \nb y_q)^{1/3}$. We fix the vector interaction parameter $a = 0.2 \rm ~fm^{-2}$ and the bag constant $B^{1/4} = 180$ MeV.

\subsection{Leptons}

The smallness of the electromagnetic fine structure constant $\alpha \simeq 1/137$, means that leptons can be treated as non-interacting, relativistic particles and therefore
\beq
\ep_L &=&\frac{1}{8\pi^2\hbar^3}\sum_l\left\{k_{Fl}(k_{Fl}^2+m_l^2)^{1/2}(2k_{Fl}^2 + m_l^2)\right.  \nn \\
      &-& \left. m_l^4\ln\left[\frac{k_{Fl}+(k_{Fl}^2 + m_l^2)^{1/2}}{m_l}\right]\right\}  \\
\mu_l &=& (k_{Fl}^2 + m_l^2)^{1/2} \\
p_L &=& \nb \sum_l y_l \mu_l - \ep_L \\
k_{Fl} &=& (3\pi^2\hbar^3 \nb y_l)^{1/3} ~~;~~ l=e,\mu~. 
\eeq

At low baryon densities only electrons are present in the system. Muons appear at a density $\nb$
such that the condition $\mu_e -m_{\mu} = 0$ is met. 

\subsection{Crossover matter}
\label{sec:KWG}
For the calculation of crossover-matter properties we rely on the Kapusta-Welle (KW) ~\cite{Kapusta:2021ney,2021PhRvD.104.l23032C} framework in the context of which the pressure is given by
\be
p_B = (1-S)p_H + S \, p_Q  ~.   \label{eqPB}
\ee
$p_H$ and $p_Q$ are the hadron and quark pure-phase pressures respectively, and the switch function
\be
S = \exp\left[-\left(\frac{\mu_0}{\mu}\right)^4\right]
\ee
gives the fraction of quark matter to the total baryonic matter when both quarks and nucleons are present. $\mu$ is the average nucleonic chemical potential
\be
\mu = \frac{n_n \mu_n + n_p \mu_p}{n_n + n_p}~\,,
\ee
and $\mu_0$ a typical energy scale for the crossover (here, $\mu_0 = 1.8$ GeV).

Applying the grand-canonical expression $n_i=\left.\frac{\p p}{\p \mu_i}\right|_{\mu_j}$ to \Eqn{eqPB} we find 
\beq
n_h^* &=& n_h\left[1-S\left(1-\frac{4\mu_0^4}{\mu^5}\frac{p_Q-p_H}{n_n+n_p}\right)\right]  \label{eqnh}  \\
n_q^* &=&  S\,n_q \,. \label{qno}
\eeq
for the crossover-matter densities (starred) of nucleons and quarks. Above, the unstarred densities are those of the pure phases. For leptons this distinction is irrelevant. 

Finally, the energy density $\ep_B$ is given by 
\be
\ep_B = -p_B + \sum_{i=n,p,u,d,s}n_i^*\mu_i  ~.
\ee
The chemical potentials in crossover matter are (functionally) the same as in the pure phases.

\subsubsection{Neutron star matter}

For a proper description of neutron-star matter that consists of nucleons, leptons and quarks, the previously unconstrained system must be subjected to the conditions of strong
\be
\mu_n = 2\mu_d + \mu_u  ~~;~~\mu_p = 2\mu_u + \mu_d 
\ee
and weak equilibrium
\be
\mu_n = \mu_p + \mu_e  ~~;~~\mu_e = \mu_\mu   ~~;~~ \mu_d = \mu_s
\ee
as well as to charge neutrality
\be
n_p^*+(2n_u^*-n_d^*-n_s^*)/3-(n_e+n_{\mu})=0 
\ee
and baryon number conservation
\be
n_n^*+n_p^*+(n_u^*+n_d^*+n_s^*)/3-\nb=0 ~.
\label{barno}
\ee
These conditions eliminate the particle fractions in favor of the total baryon density: 
\be
y_i \rightarrow y_{i,\beta}(\nb)  ~~;~~ i=n,p,u,d,s,e,\mu
\ee

\subsubsection{Sound speeds}

The total pressure and energy density in the crossover region are 
\beq
p &=& p_B + p_e + p_{\mu}  \\
\ep &=& \ep_B + \ep_e + \ep_{\mu}  ~.
\eeq
Using these, the adiabatic speed of sound is obtained by first calculating the expression
\be
\csqad(\nb,y_i) = \left.\frac{\p p}{\p \nb}\right|_{y_i}
\left(\left.\frac{\p \ep}{\p \nb}\right|_{y_i}\right)^{-1}
\ee
and then evaluating it in $\beta$-equilibrium
\be
c_{\rm{ad},\beta}^2(\nb) = \csqad[\nb,y_{i,\beta}(\nb)]~.
\ee
The equilibrium sound speed is given by the total derivatives of the pressure and the energy density  with respect to the baryon density after the enforcement of $\beta$-equilibrium,
\be
\csqeq = \frac{dp_{\beta}}{d\nb}\left(\frac{d\ep_{\beta}}{d\nb}\right)^{-1}~.
\ee
%

\subsection{Gibbs construction}

As in the crossover matter case, all thermodynamic quantities are expressed in terms of the total baryon density $\nb$, and the individual particle fractions $y_n$,~$y_p$,~$y_e$,~$y_{\mu}$,~$y_u$,~$y_d$,~$y_s$ which are, at this point, independent variables. The Gibbs construction itself, introduces another independent variable, $\chi$, which is the volume fraction of quarks in the mixed phase of a soft first-order transition such that 
\be
\ep_B = (1-\chi)\ep_H + \chi \ep_Q ~;
\ee
that is, the mixed phase is defined by the condition $0\le \chi \le 1$.

Afterwards, the conditions for mechanical, strong, and weak equilibrium, charge neutrality, and baryon number conservation are applied:
\beq
&&p_H = p_Q ~;~ \mu_n = 2\mu_d + \mu_u ~;~ \mu_p = 2\mu_u + \mu_d  \\ \nn \\
&&\mu_n = \mu_p + \mu_e ~;~ \mu_e = \mu_\mu ~;~\mu_d = \mu_s  \\ \nn \\
&&3 (1-\chi) y_p+\chi(2y_u-y_d-y_s)-3(y_e+y_{\mu}) = 0   \\  \nn \\
&&3 (1-\chi)(y_n+y_p)+\chi(y_u+y_d+y_s)-3 = 0 \,.
\eeq 

Solving these equations eliminates the $y_i$ and $\chi$ in favor of $\nb$. Thus the state variables become functions of only $\nb$ according to the rule
\[ Q(\nb,y_i,y_j,...) \rightarrow Q[\nb,y_i(\nb),y_j(\nb),...] = Q(\nb)~.\] \\
Then, the thermodynamics of the mixed $(^*)$ phase are:
\beq
\ep^* &=& (1-\chi)\ep_H + \chi \ep_Q + \ep_L  \\
p^* &=& p_H+ p_L = p_Q+ p_L \nn \\
    &=& (1-\chi) p_H + \chi p_Q + p_L  \\
\mu_h^* &=& \mu_h ~~;~~ \mu_q^* = \mu_q   \\
y_h^* &=& (1-\chi)y_h  ~~;~~ y_q^* = \chi y_q \,.
\label{qnoG}
\eeq
Quantities corresponding to leptons are not affected by the ratio of the two baryonic components in the mixed phase.

The sound speeds are obtained following the prescription outlined in the previous section.


\section{Results and Discussion}
\label{results}

\begin{table}[htbp!]
\caption{Neutron star properties for the EOSs used in this work. } 
\begin{center} 
\begin{tabular}{ccccc}
\hline
\hline
Model       & ~$L$ (MeV)  &   $M_{max}$ ($M_{\odot}$)    & $R_{max}$ (km)  & $R_{1.4}$ (km)  \\ \hline
            &  40     &  2.09   &  10.3   &  11.8      \\
  ZL        &  55     &  2.11   &  11.0   &  12.6      \\
            &  70     &  2.10   &  11.2   &  13.1      \\
\hline 
            &  40     &  2.03   &  10.6   &  11.8      \\
  KW        &  55     &  2.01   &  11.3   &  12.6      \\
            &  70     &  2.01   &  11.7   &  13.2      \\
\hline
            &  40     &  2.04   &  10.6   &  11.8      \\
  Gibbs     &  55     &  2.00   &  10.9   &  12.6      \\
            &  70     &  1.98   &  10.9   &  13.2      \\
\hline \hline
\end{tabular}
\end{center}
\label{tab:NSprops}
\end{table}

\begin{figure}[htbp!]
    \includegraphics[width=\linewidth]{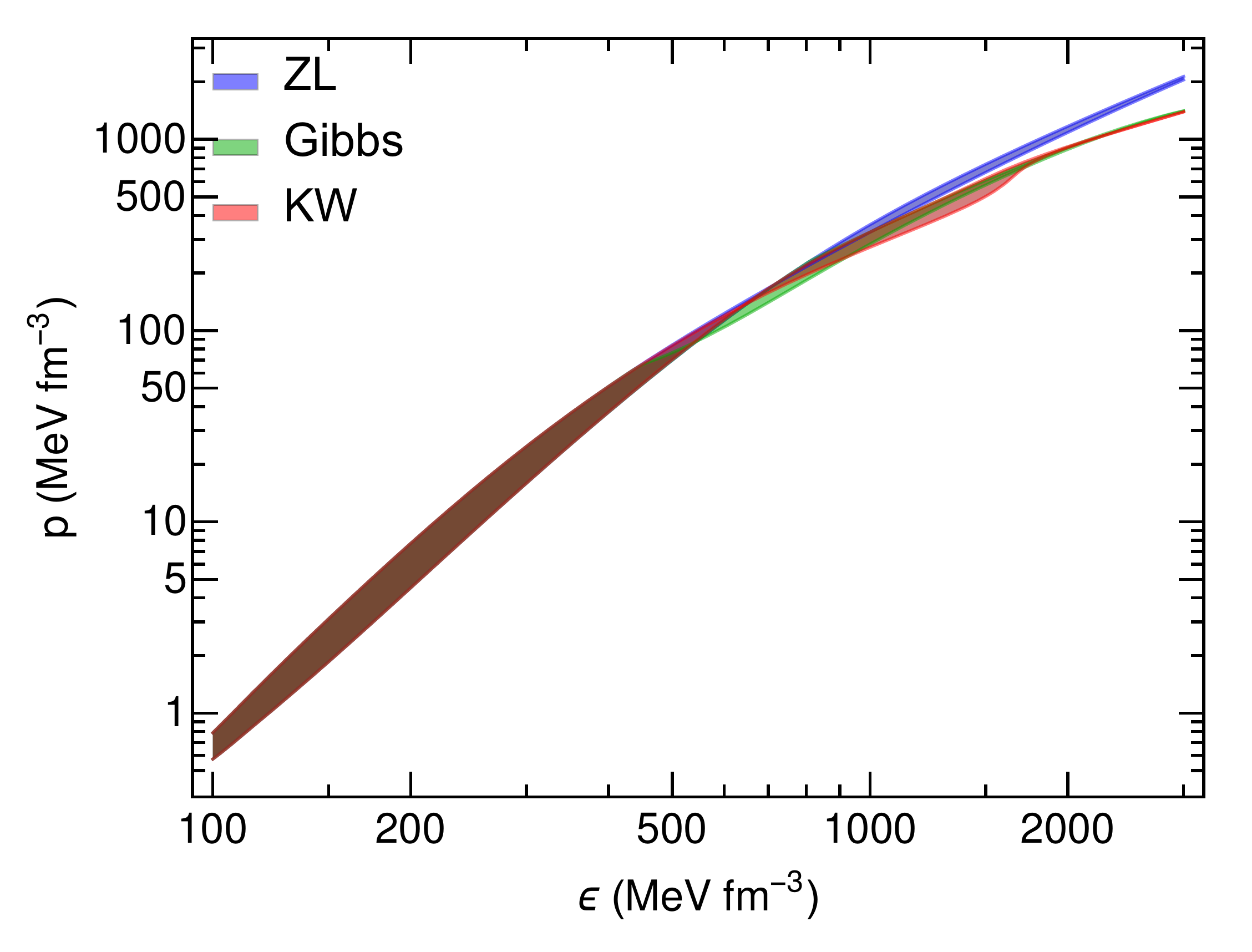}
    \caption{Range of EOSs displayed as pressure $p$ versus energy density $\varepsilon$. The composition of the three models used are: for ZL, nucleons and leptons, for Gibbs, nucleons, quarks, and leptons in a soft first-order phase transition description, and for KW, the same as for Gibbs, but in a cross-over description.}
    \label{fig:pvse}
\end{figure}

Fig. \ref{fig:pvse} is a representation of the EOSs used in this work in the pressure vs. energy-density plane. The three classes of EOSs are identical at low and intermediate energies being that quark contributions are 0 for ZL and Gibbs and vanishingly small for KW. The width of the band is a consequence of the variation of the slope of the symmetry energy $L$ in the range 40 - 70 MeV. At higher energies the importance of $L$ diminishes; here, the differences between curves are due to the presence (Gibbs and KW) or absence (ZL) of quarks and, to a lesser extent, the manner in which matter containing quarks is reached (soft first-order transition or crossover). All EOSs produce NSs consistent with the 2 $M_{\odot}$ observations; however, those corresponding to small $L$'s lead to radii that are outside the 1-$\sigma$ constraints of recent studies (e.g. \cite{Legred:2021hdx}).

\begin{figure}[htbp!]
    \includegraphics[width=\linewidth]{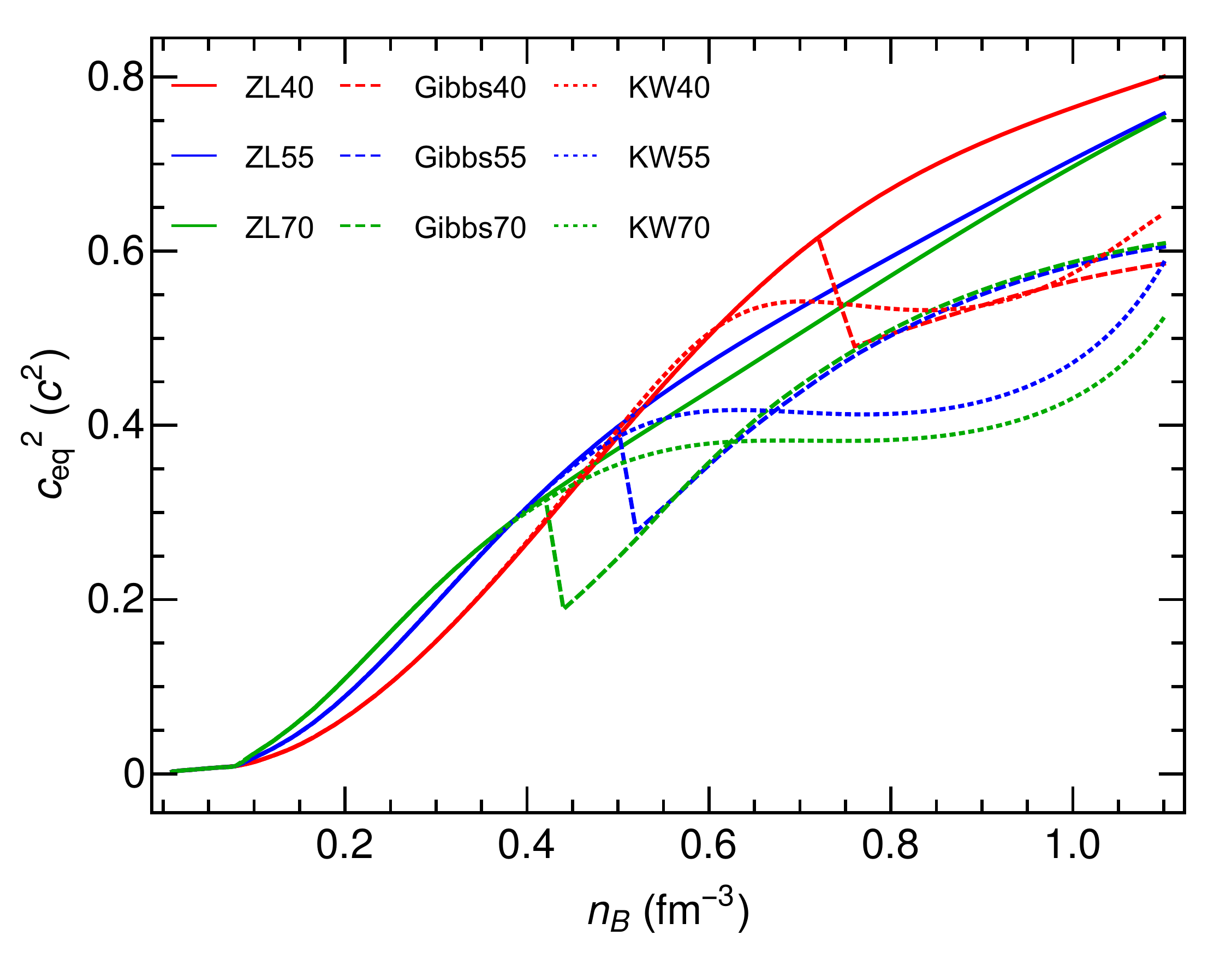}
    \caption{Squared equilibrium sound speeds vs baryon density for the models indicated in the inset. See also text for more details.}
    \label{fig:ceq}
\end{figure}

\begin{figure}[htbp!]
    \includegraphics[width=\linewidth]{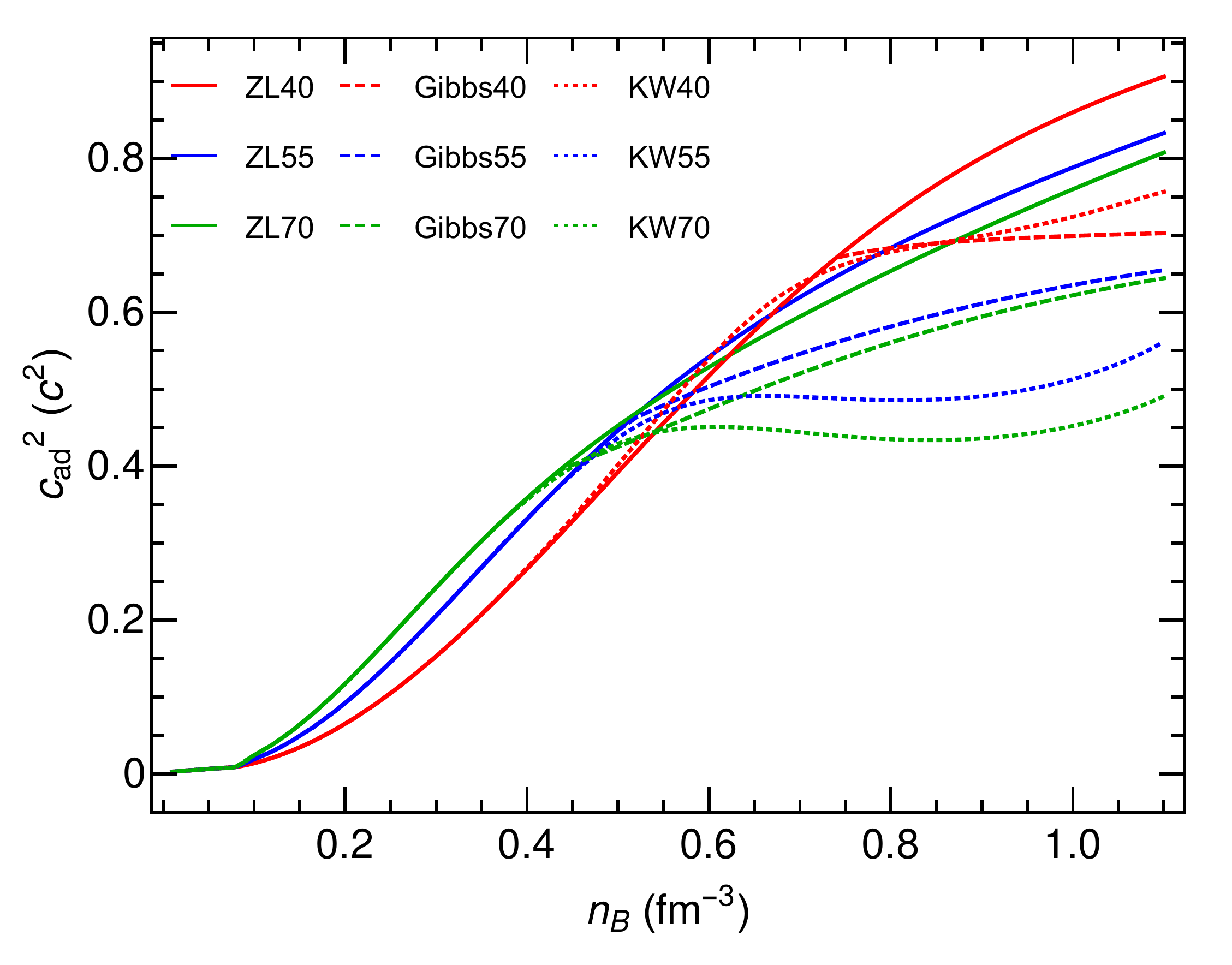}
    \caption{Squared adiabatic sound speeds vs baryon density for the same models as in Fig. \ref{fig:ceq}.}
    \label{fig:cad}
\end{figure}

The equilibrium  and adiabatic squared sound speeds vs baryon density for the three models are shown in Figs. \ref{fig:ceq} and \ref{fig:cad}, respectively, for low (40 MeV - red), intermediate (55 MeV - blue) and large (70 MeV - green) values of $L$. The curves for the purely-nucleonic model (ZL) grow monotonically and even exceed 1; albeit at densities beyond those reached in the cores of the maximum-mass NSs. Note, however, that violation of causality is not a pathology of the ZL functional but, instead, the result of our choosing a large $\gamma_1$. On the other hand, the appearance or enhancement of quarks slows down and/or reverses the growth in sound-speed (smoothly for KW and discontinuously for Gibbs).

\begin{figure}[htbp!]
    \includegraphics[width=\linewidth]{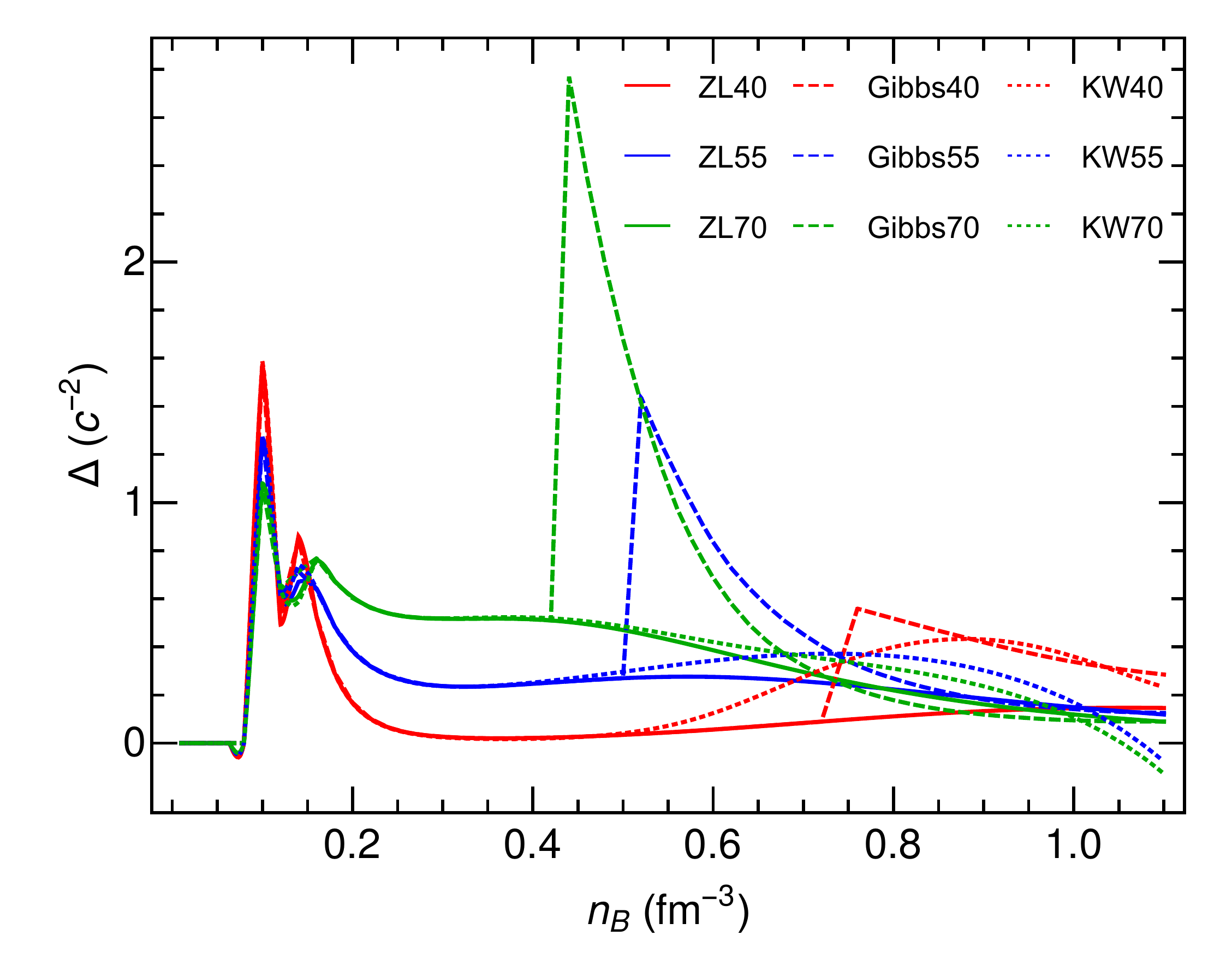}
    \caption{Difference of squared inverse sound speeds, $1/c_{\rm eq}^2 - 1/c_{\rm ad }^2$ for the same models as in Fig. \ref{fig:ceq}.}
    \label{fig:deltac}
\end{figure}

Fig. \ref{fig:deltac} shows the difference of the inverses of the two sound speeds (squared). All three models contain two sharp peaks at low densities which are due to the nuclear liquid-gas phase transition ($\sim$ 0.1 fm$^{-3}$) and to the muon onset ($\sim$ 0.15 fm$^{-3}$). The Gibbs model also exhibits sharp peaks at intermediate densities as a result of deconfinement; the height(location) of these peaks is (inversely)proportional to $L$. Less pronounced, broader peaks occur in the case of KW as well.

\begin{figure}[htbp!]
    \includegraphics[width=\linewidth]{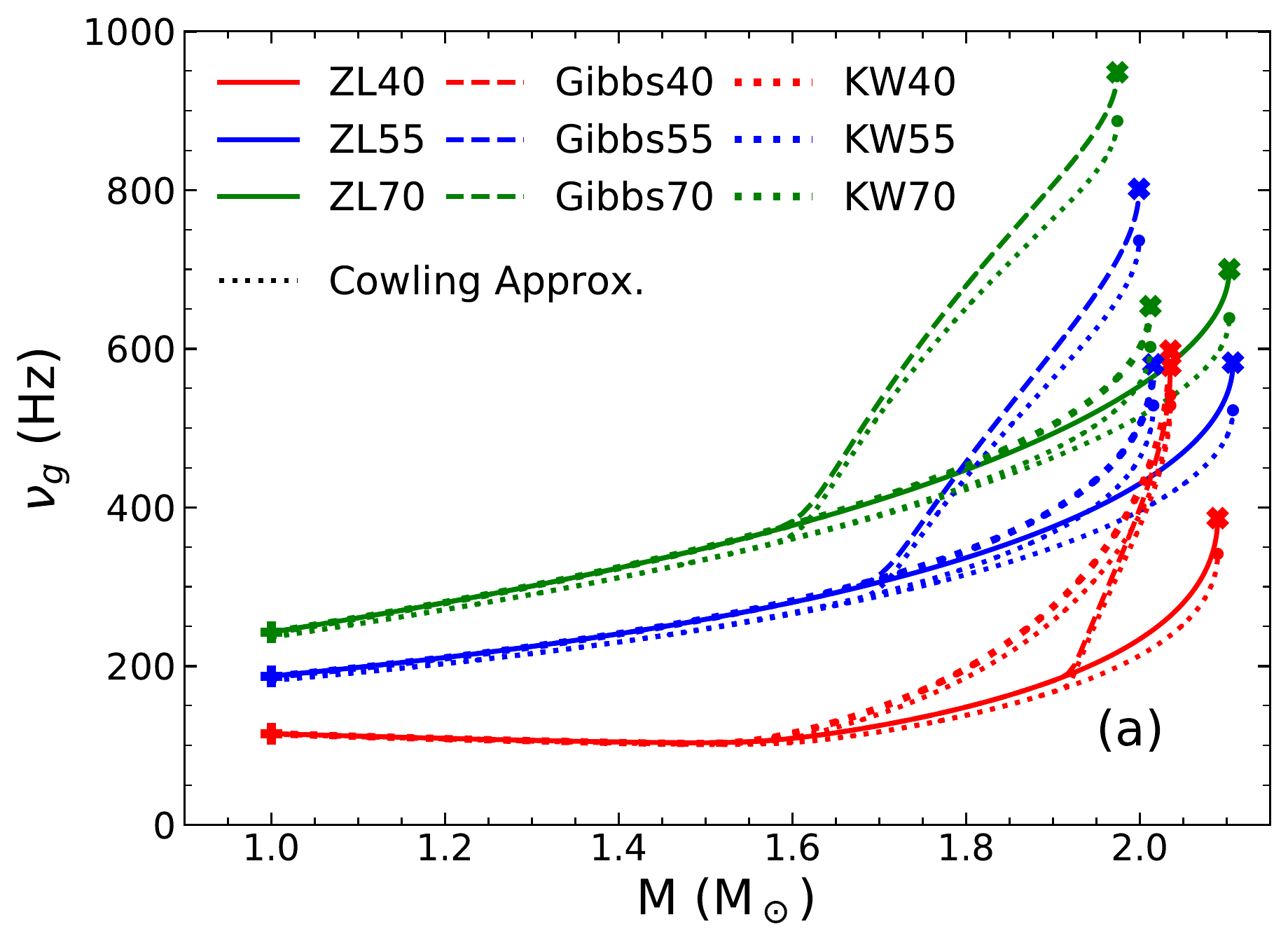}
    \includegraphics[width=\linewidth]{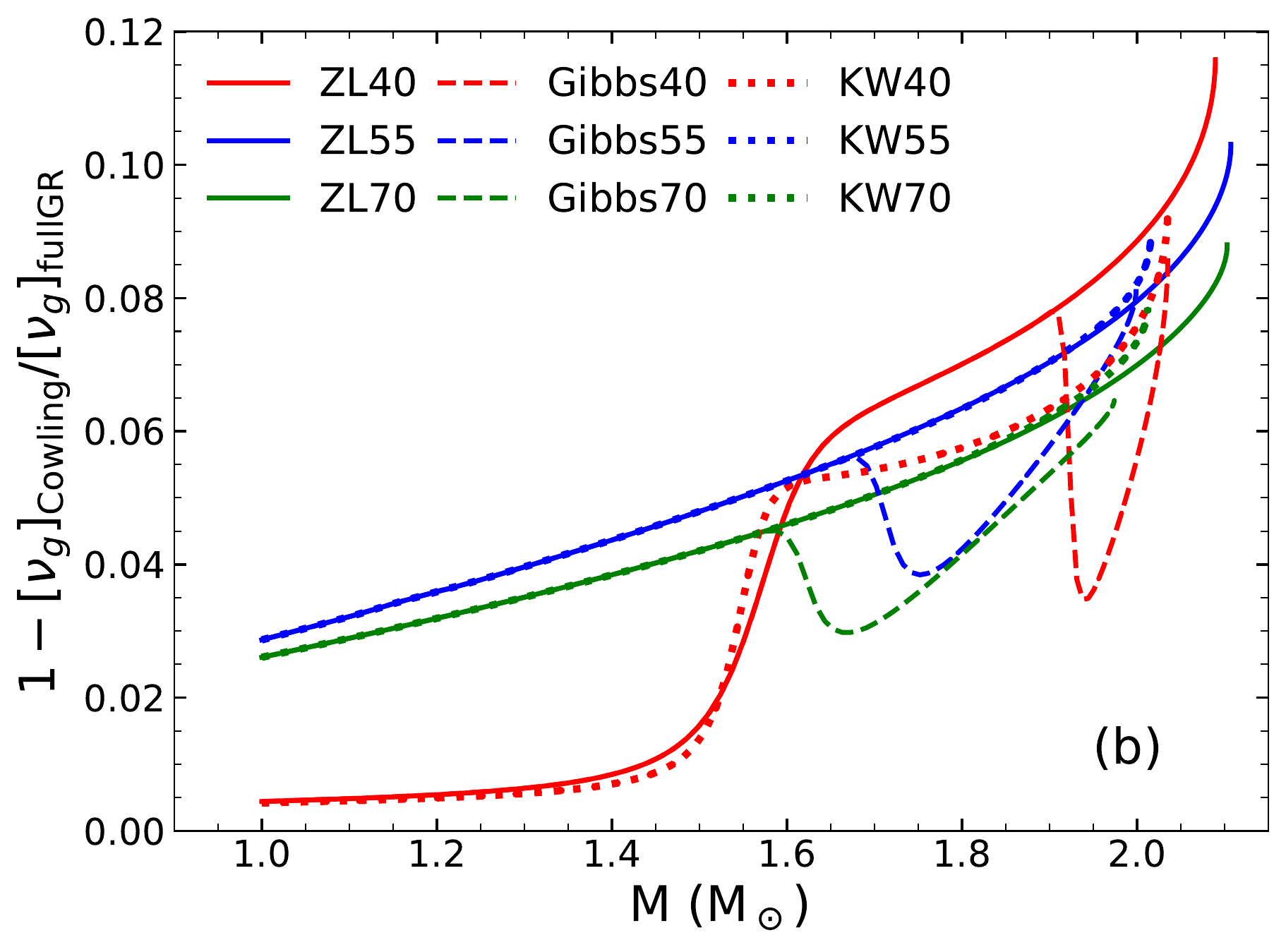}
    \caption{(a) Comparison of general relativistic and Cowling approximation (dotted curves) $g$-mode frequencies vs neutron star mass for the various EOSs considered in this work. (b) Relative errors. All curves in (a) and (b) terminate at their maximum masses.}
    \label{fig:fullGR_vs_cowling}
\end{figure}

The $g$-mode frequencies of these 9 parametrizations are shown in Figure \ref{fig:fullGR_vs_cowling} (a) for full GR calculations (solid curves)  and those using the Cowling approximation (dashed curves). The Cowling approximation generates $g$-mode frequency curves that are qualitatively similar to those  with the full GR metric perturbations. Quantitative differences do exist, however, with the Cowling approximation generally underestimating the $g$-mode frequencies by a few \% to about 10 \%.

Table \ref{tab:NSgmode} records the $g$-mode frequencies of NSs with 1M$_\odot$, 1.4M$_\odot$ and M$_{max}$ for various EOSs. Previous studies~\cite{Lai:1998yc,Sotani:2001bb} have suggested that the Cowling approximation introduces an error of less than 5\% for the $g$-mode frequency. However, our calculation suggests this is only true for NSs with low masses, M$\lesssim 1.6$ M$_\odot$. As shown in  Fig. \ref{fig:fullGR_vs_cowling} (b), the deviation increases with the NS mass, reaching $\approx 10$\% for NSs close to their maximum mass.

\begin{table}[htbp!]
\caption{$g$-mode frequency (Hz) of typical NSs for the EOSs in Table \ref{tab:NSprops} without(with) Cowling approximation.} 
\begin{center} 
\begin{tabular}{cccc}
\hline
\hline
Model       & 1 M$_\odot$ & 1.4 M$_\odot$ &   $M_{max}$\\ 
\hline
  & 114.8 (114.3) & 104.2 (103.4) & 386.4 (341.4)\\
ZL & 187.3 (181.9) & 240.6 (230.1) & 582.5 (522.4)\\
 & 242.6 (236.3) & 323.6 (311.2) & 700.2 (638.6)\\
 \hline
 & 114.8 (114.3) & 104.2 (103.4) & 578.4 (528.4)\\
Gibbs & 187.3 (181.9) & 240.6 (230.1) & 801.5 (736.3)\\
 & 242.6 (236.3) & 323.6 (311.2) & 948.4 (886.9)\\
  \hline
 & 114.6 (114.1) & 103.6 (102.9) & 597.6 (541.4)\\
KW & 186.9 (181.6) & 240.7 (230.2) & 580.3 (528.4)\\
 & 242.6 (236.3) & 324.0 (311.5) & 653.6 (602.2)\\
 \hline \hline
\end{tabular}
\end{center}
\label{tab:NSgmode}
\end{table}

\begin{figure}[htbp!]
    \includegraphics[width=\linewidth]{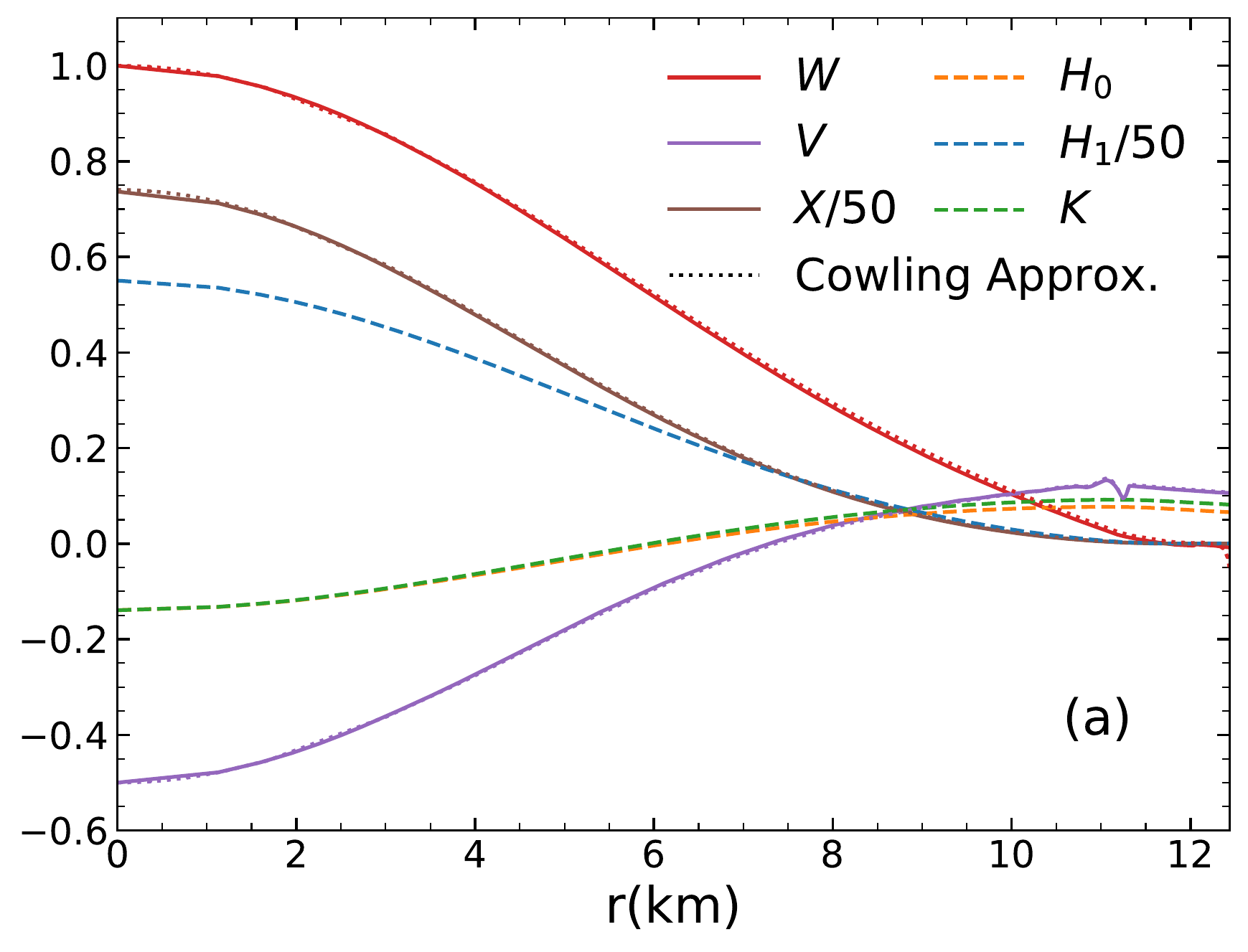}
    \includegraphics[width=\linewidth]{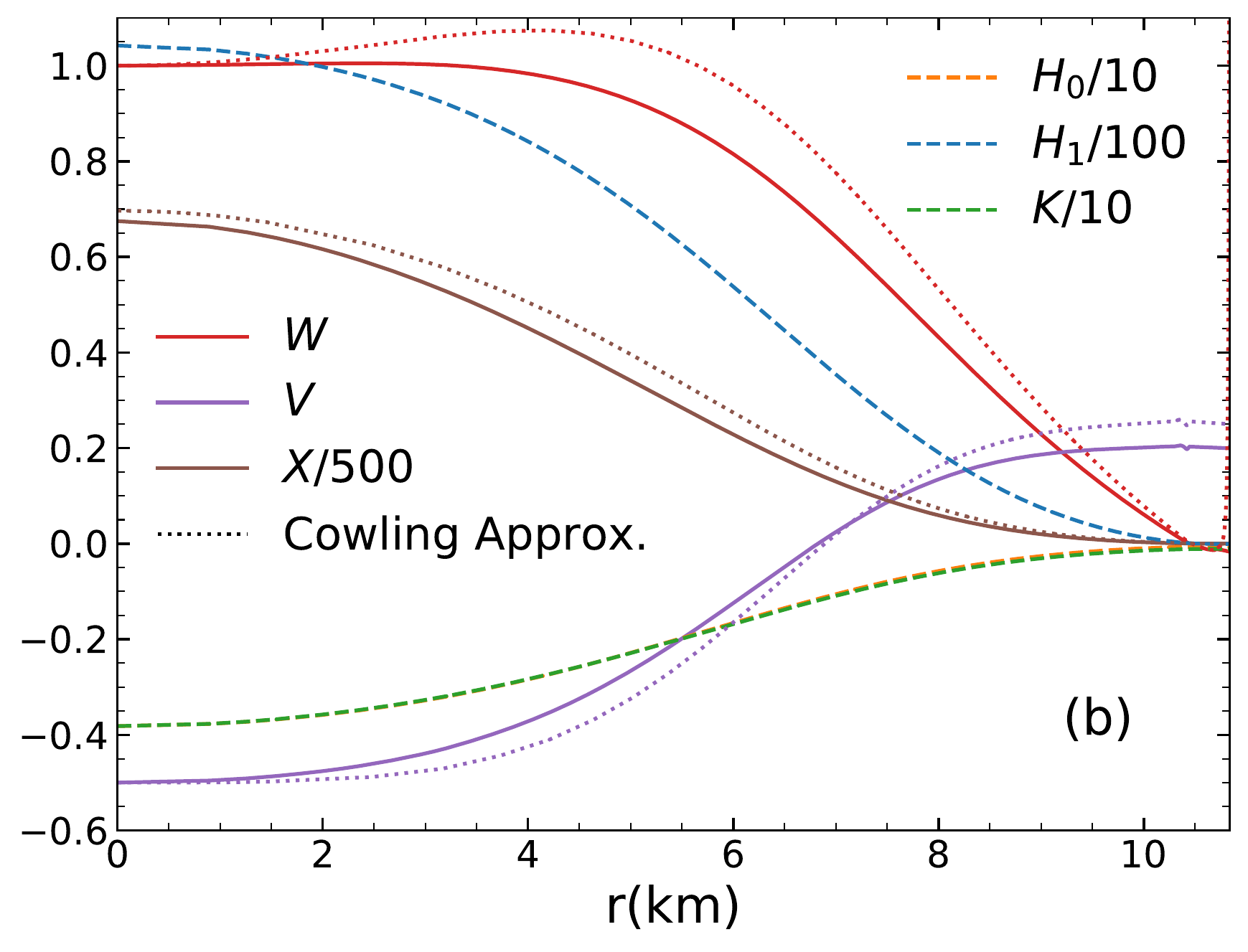}
    \caption{$g$-mode fluid perturbation functions' radial profile of general relativistic (solid curves) and Cowling approximation (dotted curves) as well as metric perturbation functions' radial profile (dashed curves) with ZL55 EOS for (a) 1.4 M$_\odot$ NS and (b) 2.11 M$_\odot$(maximum mass) NS. $W$ and $V$ are dimensionless, while $H_0$, $H_1$ and $K$ are in units of empirical energy density at saturation $\varepsilon_s=152.55$ MeV fm$^{-3}$ and $X$ is in units of $\varepsilon_s^2$.
    \label{fig:fullGR_vs_cowling_perturbation_ZLA}}
\end{figure}

The fact that 
that Cowling approximation works better for low mass NSs is not surprising insofar as they have smaller gravity. Thus, the metric perturbations corresponding to perturbations of gravity are correspondingly weak. This effect is evident in low mass NSs of the ZL40 EOS where $\Delta(c^{-2})\approx 0$ for $0.2$ fm$^{-3}$ $\lesssim n_B \lesssim 0.7$ fm$^{-3}$. Therefore, the $g$-mode oscillation is supported mostly by matter close to the surface of the NS where gravity is weak, which allows the Cowling approximation to be particularly accurate, see Fig. \ref{fig:fullGR_vs_cowling} (b).
We illustrate this comparing results for a 1.4 M$_\odot$ NS with those of a 2.11 M$_\odot$ (maximum mass) NS with the ZL55 EOS.
Figure \ref{fig:fullGR_vs_cowling_perturbation_ZLA} shows the fluid and metric perturbation functions for the above two cases with (dashed curves) and without (solid curves) the Cowling approximation. 
In the case of the 1.4 M$_\odot$ NS, the Cowling approximation produces a nearly identical fluid perturbation profile as GR with the full metric perturbation. As a result, the characteristic frequency decreases from 240.6 Hz to 230.1 Hz with an error of less than 5\%. The small kink in the metric function $V$ close to the surface in panel (a) arises due to the core-crust transition which is negligible in panel (b). In the case of the 2.11 M$_\odot$ NS, the average magnitudes of the metric perturbation are about an order of magnitude larger than that for the 1.4 M$_\odot$ NS. In addition, the fluid perturbation function with the Cowling approximation deviates from GR significantly. The characteristic frequency decreases from 582.5 Hz to 522.4 Hz with an error of about 10\% when the Cowling approximation is used.

\begin{figure}[htbp!]
    \includegraphics[width=\linewidth]{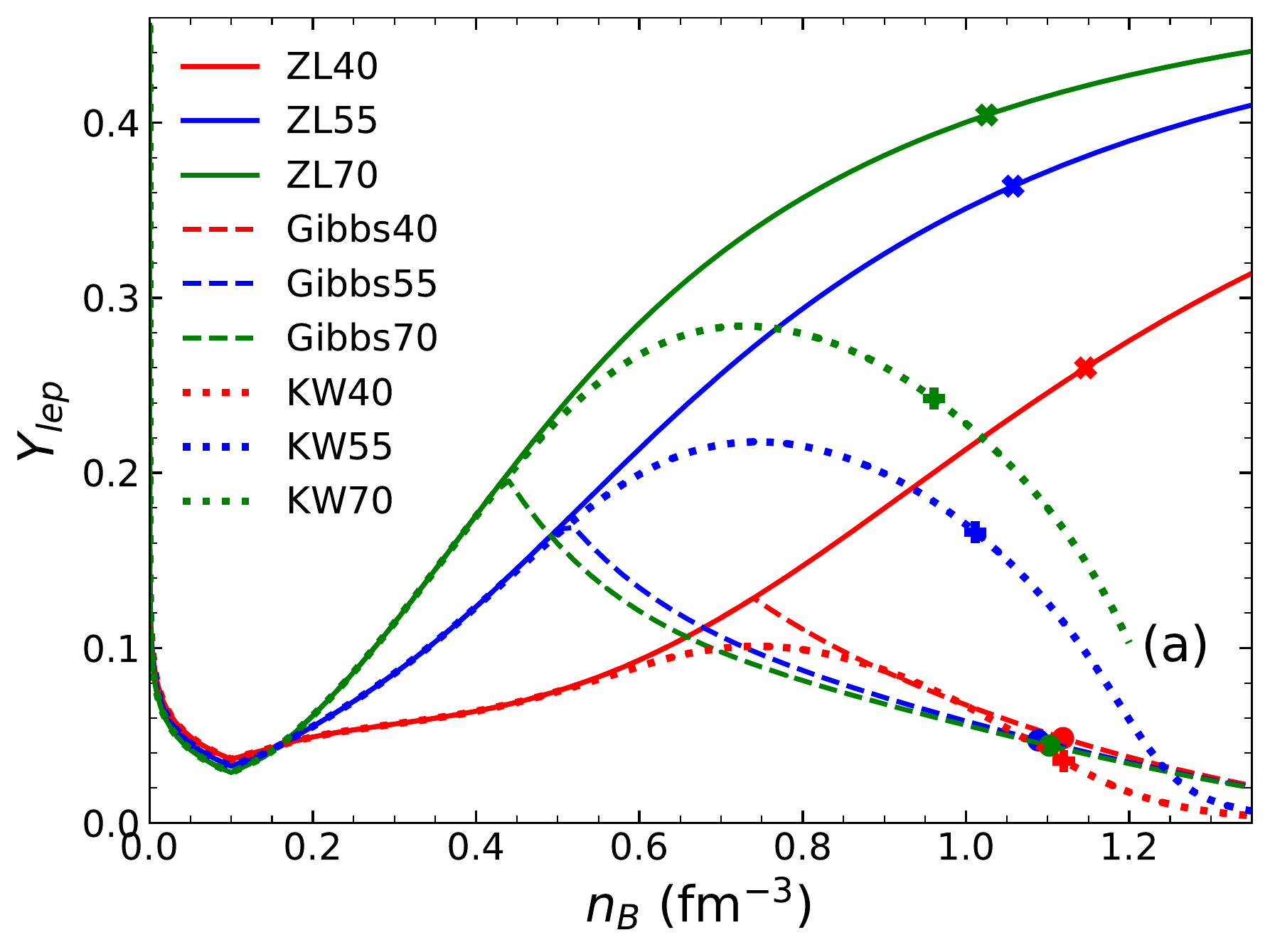}
    \includegraphics[width=\linewidth]{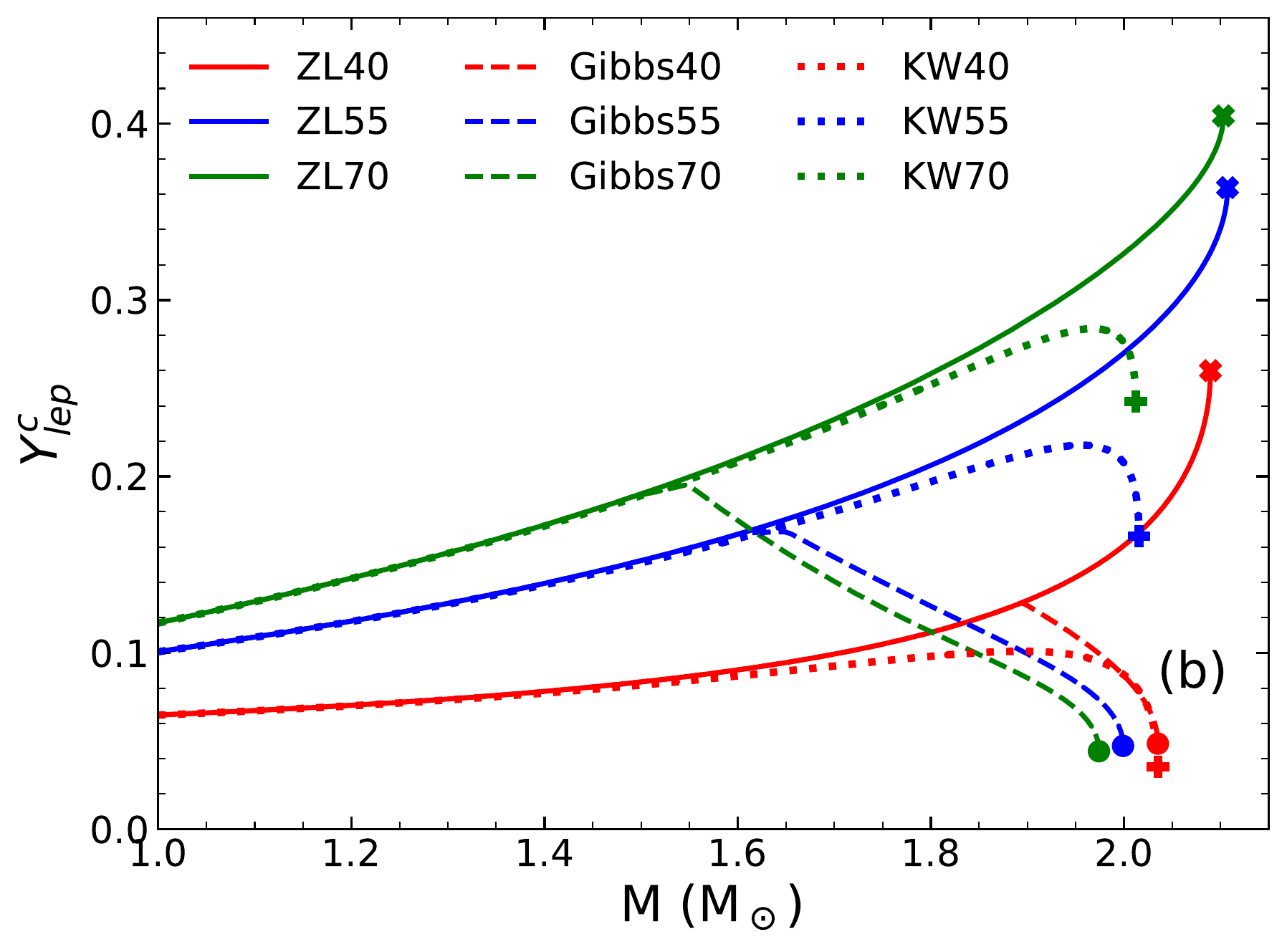}
    \caption{(a) Lepton number fraction as a function of baryon number density. (b) Lepton number fraction at the center of NS as a function of NS mass. Markers on the curves indicate the maximum mass configurations of the corresponding EOS. }
    \label{fig:Ylep_nB_mass}
\end{figure}
\begin{figure}[htbp!]
    \includegraphics[width=\linewidth]{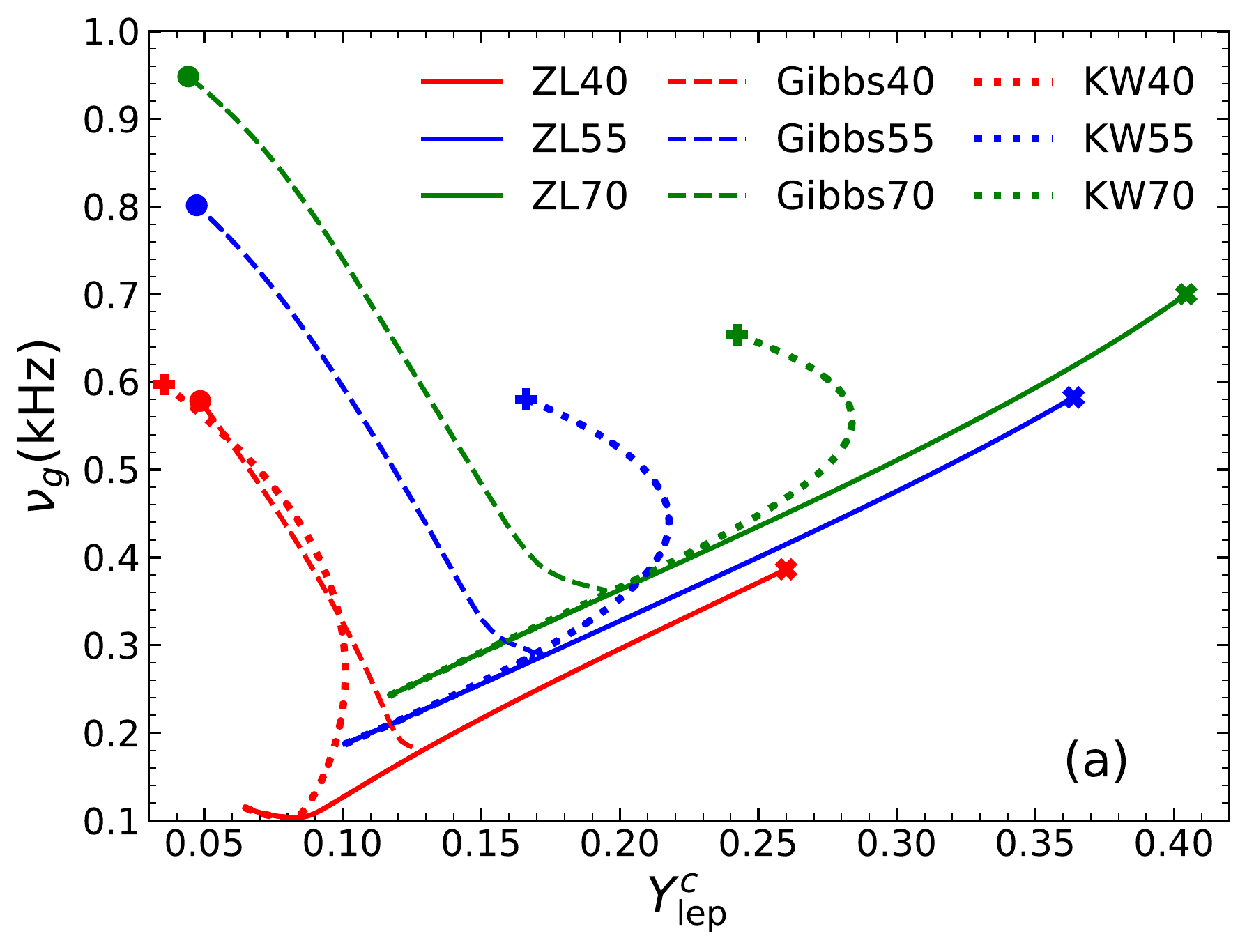}
    \includegraphics[width=\linewidth]{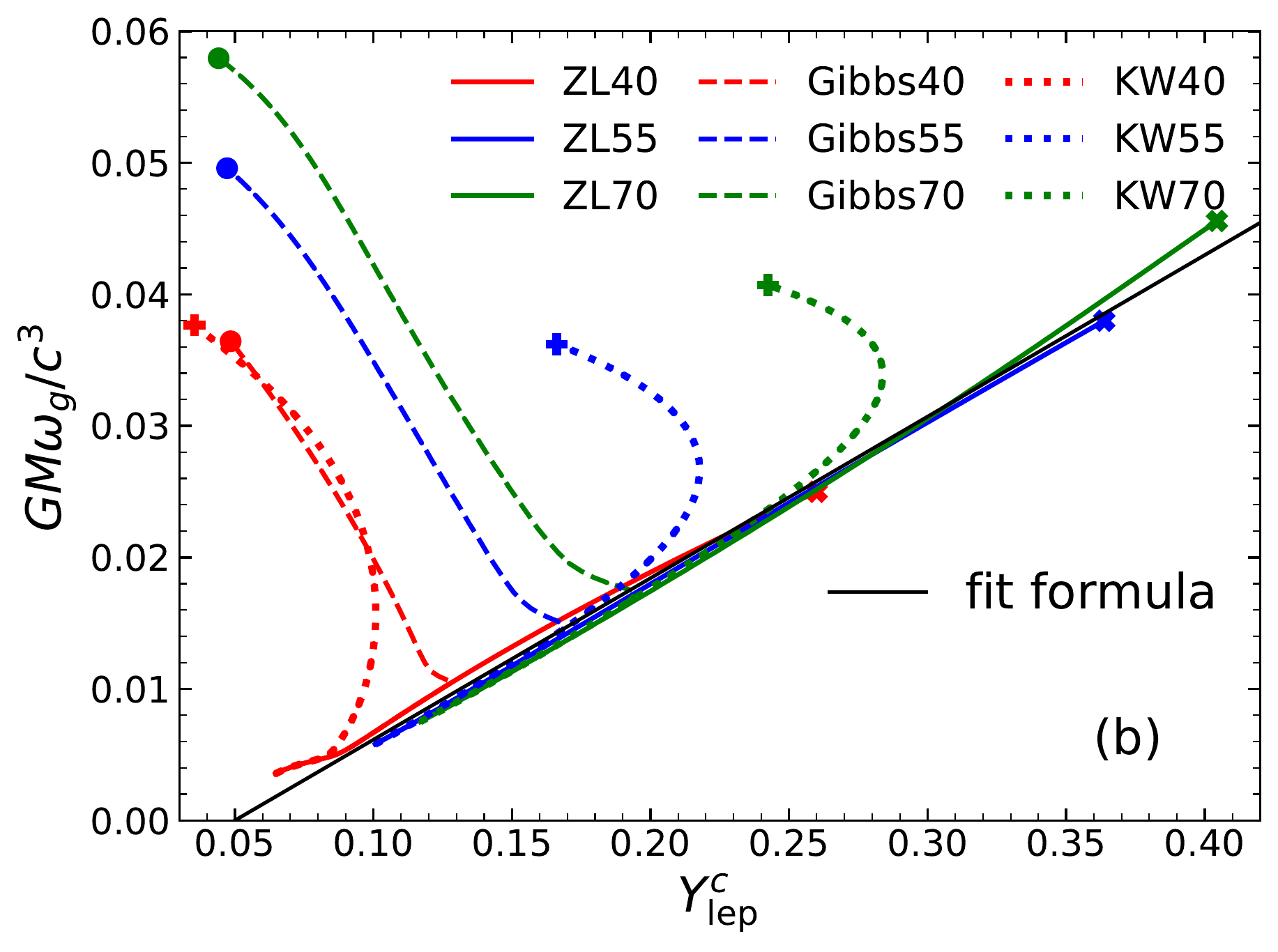}
    \caption{(a) Comparison of the $g$-mode frequencies vs lepton fraction for the various EOSs with and without quarks. (b) Same as (a) but for the dimensionless quantity $GM\omega_g/c^3$ vs lepton fraction.}
    \label{fig:cowling_Ylep}
\end{figure}

One of the goals 
of this paper is to investigate the existence of a scaling relation involving the $g$-mode frequency and an internal attribute of the NS. Previous studies have shown that the  $f$-mode universally correlates with compactness~\cite{andersson1998towards} and moment of inertia~\cite{lau2010inferring}, and the $p$-mode correlates with mean density~\cite{kokkotas2001inverse}, the discontinuous $g$-mode correlates with the density discontinuity and transition density~\cite{finn1987g,mcdermott1990density,miniutti2003non,zhao2022universal}. 
Since the restoring force for the $g$-mode is the tendency towards chemical equilibrium, the associated  frequency should be related to the frequency of the local chemical oscillation in matter which is proportional to the difference between the equilibrium and adiabatic sound speeds. This difference depends strongly on the lepton fraction in equilibrium in the case of NS matter. Thus, we investigate the role of the lepton fraction in the scaling relation regarding $g$-mode frequency.

The lepton fraction $Y_{\rm lep}=y_e+y_\mu$ as a function of baryon number density is determined by the EOS.  Figure  \ref{fig:Ylep_nB_mass} (a) shows the trends for the various EOSs considered in this work. The markers on these curves indicate the maximum mass configurations. The monotonic increase of $Y_{\rm lep}$ with $n_B$ is due to a similar monotonic increase of the symmerty energy for ZL EOSs with nucleons and leptons only. 
In contrast, the lepton fractions of hybrid EOSs deviate from those of nucleonic EOSs at the onset of quarks.  The downward trends with $n_B$ are due to the fact that the charged quarks render the fractions of leptons to be diminished in satisfying charge neutrality. 
In the case of Gibbs construction, the quark-hadron transition is of first-order, leading to a sharp kink in $Y_{\rm lep}$. The KW crossover treatment uses a smooth bridge between the quark and nucleonic EOSs, and results in a smooth $Y_{\rm lep}$. Hybrid NSs with both Gibbs and KW constructions approach the limit of pure quark matter at high density causing $Y_{\rm lep}=0$.

Figure \ref{fig:Ylep_nB_mass} (b) shows the 
lepton fraction at the center of the NS, $Y_{\rm lep}^c$, taken to be its characteristic lepton fraction, as a function of mass. Note that $Y_{\rm lep}^c$ follows the same trend as in Fig. \ref{fig:Ylep_nB_mass} (a), with the $x$-axis scaled. The curves become steep on the right side since the central density increases quickly toward maximum mass configurations.

The $g$-mode frequency is shown in Fig. \ref{fig:fullGR_vs_cowling} (a). Results for the nucleonic EOS (ZL) have identical shapes as in Fig. \ref{fig:Ylep_nB_mass} (b) and Fig. 
 \ref{fig:fullGR_vs_cowling} (a), indicating a strong correlation between the central lepton fraction and the $g$-mode frequency. We introduce a dimensionless $g$-mode frequency as $\Omega_g=GM\omega_g/c^3$, which universally correlates with the central lepton fraction for nucleonic NSs, see Fig (b). \ref{fig:cowling_Ylep}. This correlation is well fit with the linear form
\begin{eqnarray}
\Omega_g=1.228(Y^c-0.05) \,, \label{eq:Omega_g_universal}
\end{eqnarray}
where $Y^c=Y_{\rm lep}^c$ is a characteristic of weak equilibrium which dominates the $g$-mode in nucleonic NSs.

The hybrid NSs, however, deviate from the above correlation. Figure \ref{fig:cowling_Ylep} shows that $Y_{\rm lep}$ decreases with density after the onset of quarks, whereas the $g$-mode frequency keeps increasing. Because the quark-hadron mixture introduces additional strong equilibrium between nucleons and quarks,  we characterize the quark content of hybrid NSs by the quark number fraction $Y_{\rm qak}=(y_u+y_d+y_s)/3$.

\begin{figure}[htbp!]
    \includegraphics[width=\linewidth]{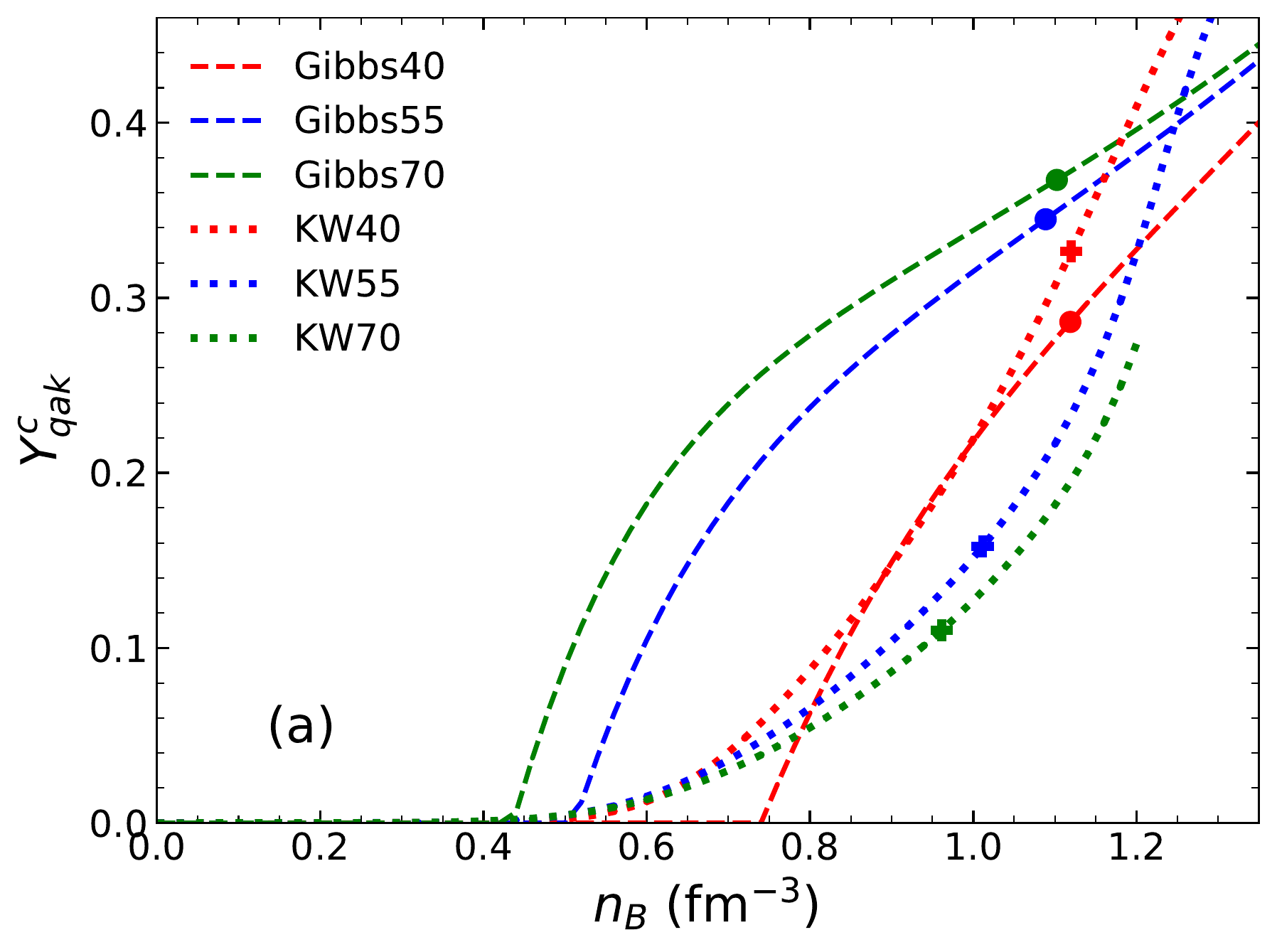}
    \includegraphics[width=\linewidth]{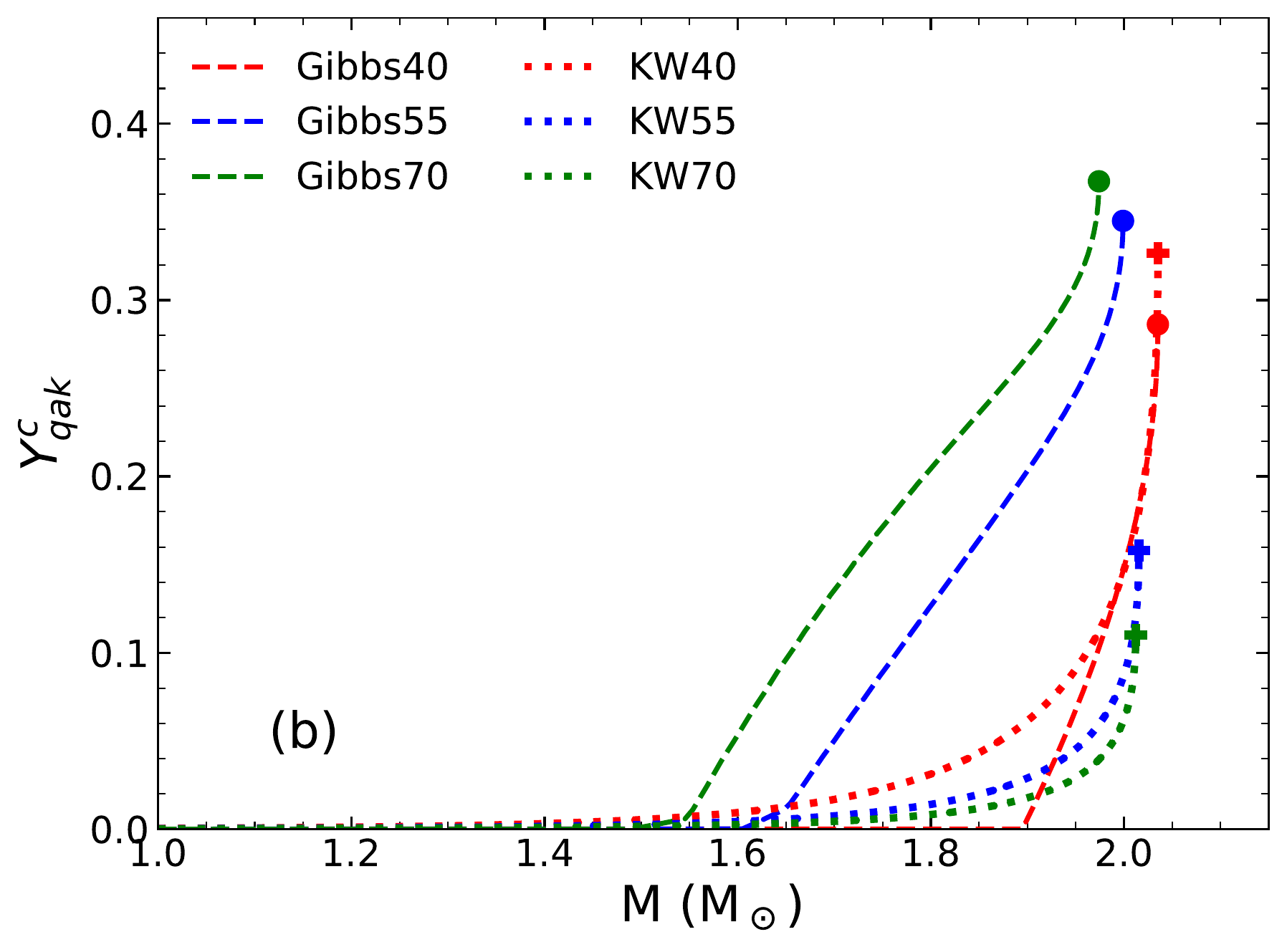}
    \caption{Same as Fig. \ref{fig:Ylep_nB_mass} but for quark fraction $Y_{\rm qak}$ in hybrid NS EOS.}
    \label{fig:Yqak_nB_mass}
\end{figure}

In order to combine the contribution from strong and weak equilibrium, we take $Y_{\rm lep}+Y_{\rm qak}$ as a new dimensionless parameter to obtain an EOS-insensitive relation for the $g$-mode frequency, as shown in Figure \ref{fig:cowling_Ylep}. Results of $Y_{\rm qak}$ for hybrid NS EOSs are shown in Fig. \ref{fig:Yqak_nB_mass}. The quark fraction increases steeply with density for Gibbs, whereas it increases slowly at low density and become steep at higher density for KW. A similar tendency can be observed in $g$-mode frequency as well, see Fig. \ref{fig:fullGR_vs_cowling} (a), indicating a strong correlation between the central quark fraction and the $g$-mode frequency.

\begin{figure}[htbp!]
 \includegraphics[width=\linewidth]{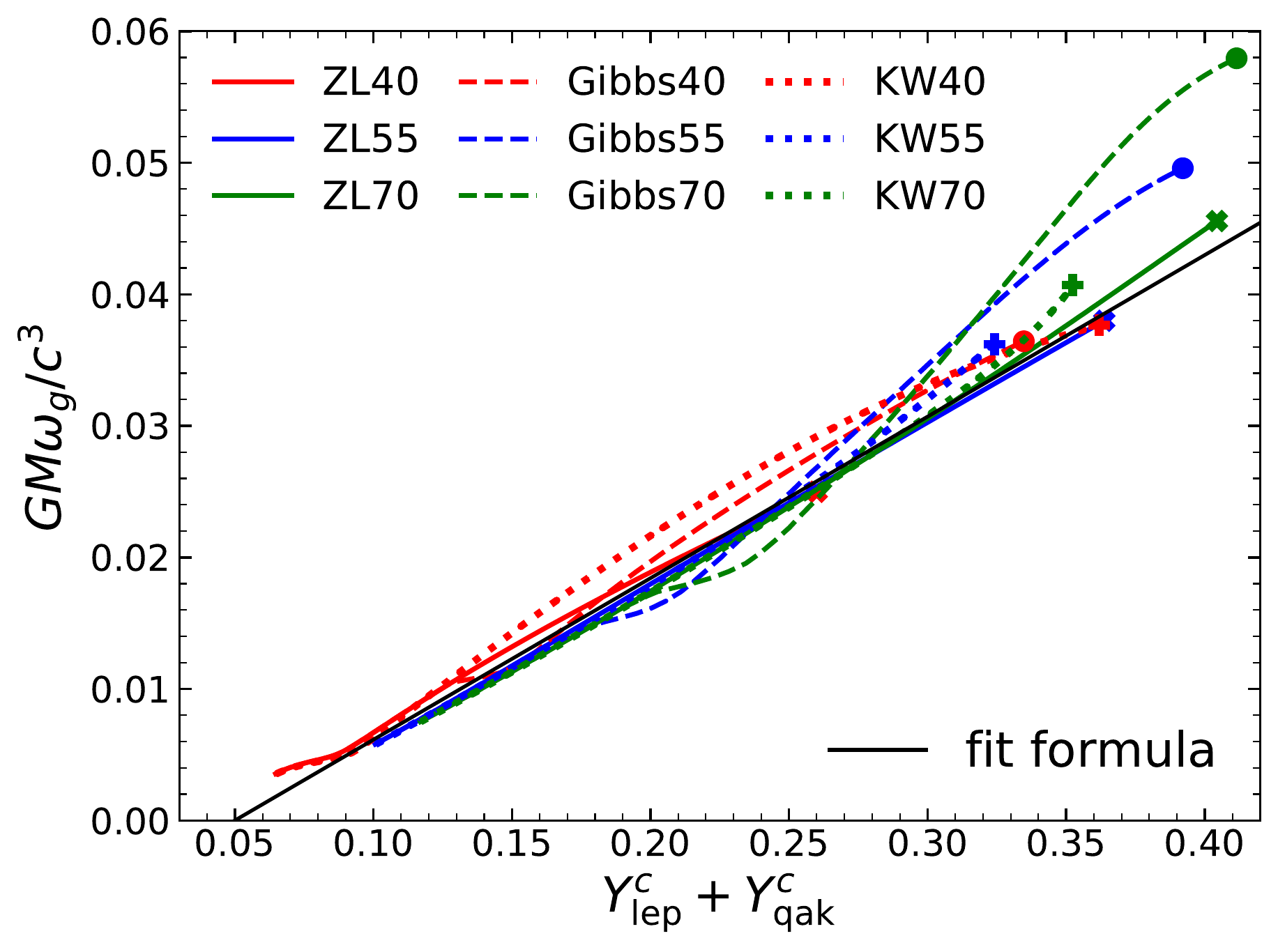}
    \caption{Same as Fig. \ref{fig:cowling_Ylep} (b) but with x-axis replaced with $Y_{\rm lep}+Y_{\rm qak}$ for hybrid NS EOS.}
    \label{fig:cowling_YlepYqak}
\end{figure}

The universal relation in Eq. (\ref{eq:Omega_g_universal}) can be used for hybrid NSs with $Y^c = Y_{\rm lep}^c+Y_{\rm qak}^c$. Given the fact that ZL, Gibbs and KW EOSs have drastically different compositions, this universal relation is adequate enough, compared with the 10\% deviation caused by the Cowling approximation. Results illustrating this relation are shown in Fig. \ref{fig:cowling_YlepYqak}.

\section{Summary and Conclusions}
\label{sec:concs}
In this work, we extended the calculation of $g$-mode oscillations in hybrid stars that we initiated within the Cowling approximation in~\cite{jaikumar2021g,2021PhRvD.104.l23032C}, to the linearized perturbation equations of general relativity. Our main objectives were to quantify the differences in the $g$-mode spectrum for nucleonic and hybrid stars between the full GR and Cowling approximation approaches, as well as to obtain new scaling relations between mode frequencies and global stellar properties. We utilized self-consistent fluid and metric perturbation equations employed in~\cite{Thorne:1967b,Lindblom:1983}, taking care to choose variables that are singularity-free inside the star and verifying their limiting forms that yield the equations of the relativistic Cowling approximation. The microscopic description of nucleonic and quark matter is identical to our previous work~\cite{2021PhRvD.104.l23032C}, allowing for a direct comparison of results between the Cowling and general relativistic framework. We also computed the damping times for $g$-modes and found them to be very long compared to merger timescales or any other dynamical timescales over which $g$-modes might be excited, indicating that these modes are long-lived. Such modes, if excited to sufficient amplitude, can act as a source of gravitational wave emission owing to the CFS instability~\cite{CFS,FS} in rotating neutron/hybrid stars, providing a link between the gravitational wave signal and the composition of the star. We have not explored dissipative effects on the $g$-mode which can considerably narrow the instability window in rotation rate and temperature~\cite{Lai:1998yc}. Different formalisms~\cite{Israel-Stewart, Noronha} have been suggested to address causality and stability of perturbations in relativistic dissipative self-gravitating fluids, which can be avenues for future work.

Our principal finding is that, for any given stellar configuration up to the maximum mass (about 2-2.25$M_{\odot}$ depending on the EOS), the fundamental $g$-mode (i.e., the one with the lowest frequency) in general relativity agrees to within $\approx$ 10\% with that obtained in the Cowling approximation, with the precise amount of deviation decreasing with decreasing stellar mass. This trend is expected due to the decreasing relevance of general relativity for lower mass neutron stars, but we find deviations to be larger than previously suggested~\cite{Sotani:2001bb,Lai:1998yc}, especially for higher mass stars. This finding holds whether the star is purely composed of nucleons, or if it contains an admixture of quarks. We therefore conclude that results for $g$-mode frequencies in the Cowling approximation are fairly robust across the range of observed neutron star masses for both nucleonic and hybrid stars.

Interestingly, the deviations between the Cowling and general relativistic results are largest for nucleonic stars, while for hybrid stars modeled by the Gibbs construction, they decrease substantially at the onset of the quark phase (threshold NS mass $\approx$ 1.8$M_{\odot}$) before rising again as the maximum mass is approached. We suspect this is due to the fact that $g$-mode frequencies are mainly determined by differences between the equilibrium and adiabatic sound speeds, which in turn depend on the input microscopic EOS rather than general relativistic effects; as the sound speed difference increases sharply when quarks appear in the Gibbs phase, their impact on the $g$-mode frequencies overwhelms any effects that stem from general relativity. This would be consistent with the clearly non-monotonic behavior in the deviation that is evident in the Gibbs construction, but is absent from the crossover models where quarks are admixed with nucleons at any density. In other words, while it is possible to find crossover models for hybrid stars where the Cowling approximation is an excellent approximation for any neutron star mass, the reliability of the approximation in the Gibbs case depends on the neutron star mass.

This is not to imply that the inclusion of metric perturbations has no effect on the $g$-mode itself. In neutron star masses of $\approx$ 2$M_{\odot}$ or larger, metric perturbation amplitudes in the star are an order of magnitude larger compared to a canonical mass of $\approx$ 1.4$M_{\odot}$, effectively decreasing the size of the fluid perturbations by up to 10\% compared to the Cowling result. Ultimately, the coupling of the fluid to the background metric is what determines the gravitational wave amplitude, and the detectability of the gravitational wave signal.

By studying trends in the $g$-mode with composition, we also found a universal relation between the (dimensionless) $g$-mode frequency and the central lepton (or lepton+quark) fraction of a purely-nucleonic (hybrid) NS. In both cases, this is understood to be a consequence of the dependence of the sound speed difference on changes in lepton and quark fractions as these particle species drive reactions that restore weak and strong equilibrium, respectively, in the perturbed fluid. While the simple universal relation presented here works remarkably well for a purely-nucleonic star, it is to be employed with caution for hybrid stars, as compositional changes can be varied and sudden, depending on how the phase transition is modeled therein. In general, the applicability of this universal relation is less sensitive to the inclusion of general relativity than it is to compositional changes in the hybrid EOS, with the Gibbs construction deviating more than the crossover models. 
Although no core oscillation modes have yet been detected from gravitational waves or electromagnetic observations, such universal relations and their relation to global stellar attributes are of practical value in constraining mode frequencies, as has been demonstrated for the case of $f$-modes~\cite{Pratten:2019sed}. One hopes that in the future, with third generation detectors like the Einstein Telescope or Cosmic Explorer, either through the direct detection of $g$-modes or its coupling to transient electromagnetic bursts~\cite{RG2}, we can have conclusive evidence about the composition of neutron star interiors.

Our results are of relevance to the late stages of a binary NS merger, particularly during the inspiral phase when the tidal field reaches resonance with either star's internal oscillation modes, resulting in energy and angular momentum transfer from matter to gravitational waves. Amongst these modes, while the $f$-mode has traditionally been the focus of study, the $g$-mode is unique in its sensitivity to composition, therefore, studying the effect of general relativity on the $g$-mode of neutron/hybrid stars is deserving of further study. The lower frequency $g$-mode could excite resonance at the earlier stage of a merger, making it much more likely to be observed in GW detectors. Building on our work presented here, incorporating effects such as rotation~\cite{Doneva, Steinhoff2021,Kruger2021} and superfluidity \cite{Gualtieri2014, Gusakov2013} will help in understanding the $g$-modes of hybrid stars more thoroughly.

\begin{acknowledgments}

T.Z and M.P. are supported by the Department of Energy, Grant No. DE-FG02-93ER40756. 
C.C. acknowledges support from the European Union's Horizon 2020 research and innovation programme under the Marie Sk\l{}odowska-Curie grant agreement No. 754496 (H2020-MSCA-COFUND-2016 FELLINI).
P.J. is supported by the U.S. National Science Foundation Grant No. PHY-1913693.

\end{acknowledgments}

\newpage

\bibliography{GRRvsCOW}
\end{document}